\documentclass[aip,
               jcp,
               preprint,
               amsmath,
               superscriptaddress,
               floatfix,
               nobibnotes,
               ]{revtex4-2}

\usepackage{amssymb}
\usepackage{amsmath}
\usepackage{mathtools}
\usepackage{latexsym}
\usepackage{amsfonts}
\usepackage[mathscr]{eucal}
\usepackage{bm}
\usepackage[dvips]{graphicx}
\usepackage{color}
\usepackage[utf8]{inputenc}
\usepackage[flushleft]{caption}
\usepackage{subfigure}
\usepackage{rotating}
\usepackage{txfonts}
\usepackage{booktabs}
\usepackage[unicode]{hyperref}
\DeclareMathAlphabet{\mathpzc}{OT1}{pzc}{m}{it}

\def\figfoot{Chem-Dyn-QuanSimu}
\def\mathbi#1{\textbf{\em #1}}

\def\mi{\textrm{i}}

\makeatletter

\makeatletter

\newcommand{\figcaption}[2]{
    \noindent {\bf Figure \ref{#1}:} #2
    \vspace{1cm}}

\begin{document}

%---------------------------------------------------
% Title, authors, and address
%---------------------------------------------------

\title{Implementation Possibility of Quantum Simulation for Quantum Molecular Dynamics}

\author{Xingyu Zhang}
\thanks{These authors contributed equally to this work.} %, where X.Z.
%developed the theory and W.G. implemented it together.}
 \affiliation{Theoretical Chemistry,
              Department of Chemistry, 
              Northwestern Polytechnical University,
              West Youyi Road 127, 710072 Xi'an,
              China}

\author{Weijia Guo}
\thanks{These authors contributed equally to this work.} % where X.Z.
%developed the theory and W.G. implemented it together.}
 \affiliation{Theoretical Chemistry,
              Department of Chemistry, 
              Northwestern Polytechnical University,
              West Youyi Road 127, 710072 Xi'an,
              China}

\author{Jinke Yu}
 \affiliation{Theoretical Chemistry,
              Department of Chemistry, 
              Northwestern Polytechnical University,
              West Youyi Road 127, 710072 Xi'an,
              China}
                            
\author{Qingyong Meng*}
\thanks{To whom correspondence should be addressed}
 \email{qingyong.meng@nwpu.edu.cn}
  \affiliation{Theoretical Chemistry,
               Department of Chemistry, 
               Northwestern Polytechnical University,
               West Youyi Road 127, 710072 Xi'an,
               China}

\date{\today}

%----------------------------------------------------------
% Abstract
%----------------------------------------------------------

\begin{abstract}

\noindent {\bf Abstract}:
In this work, we explore the implementation possibility of quantum
simulation for quantum molecular dynamics, in particular for reaction
dynamics, though several implementations have already reported through
quantum-classical mixed simulations ({\it Acc. Chem. Res.} {\bf 54}
(2021), 4229 and {\it J. Phys. Chem. Lett.} {\bf xx} (2026), XXXX).
To analyze this aspect, we examine (1) the conjugacy relation
between quantum simulator and the target molecular system, (2) the
wave function correspondence in quantum algorithm and classical
algorithm for multi-dimensional dynamics, (3) problems arisen from
real-valued classical algorithms, and finally (4) geometric phase
arisen from the separation among the degrees of freedom (DOFs). As
is well known, the aforementioned first and second points play fundamental
roles in quantum simulation of quantum many-body systems, and the third
and fourth points are theoretical issues that might introduce problems in
classical and quantum computing. In this work, we mainly focus on
the third and fourth points by analysis of the first two points by reviewing
previously reported quantum-classical mixed implementations of quantum
simulation. We also consider gauge freedom in high-dimensional quantum
molecular dynamics that has been introduced recently, and then discuss
possibility of advantages and disadvantages of quantum simulation for
molecular reaction dynamics.
{\it Quantum Computing}

\end{abstract}
\maketitle

%---------------------------------------------
% Introduction
%---------------------------------------------
\section{Introduction\label{sec:intro}}

Quantum computing offers a fundamentally efficient and natural platform
for simulating quantum many-body systems, as it directly obeys the same
quantum-mechanical laws that govern the target system. Quantum computing
for chemistry motions needs development of both quantum simulator devices
and associated quantum algorithms, in particular variational quantum
algorithms (VQAs). Both tasks are, however, challenging
\cite{cle10:1155,geo14:153,gue19:045001,cao19:10856,oll20:260511,oll20:043140,oll21:4229,bha22:015004,zhu26:xxx}.
For example, the superconducting circuits primarily focus on electronic
energies at fixed nuclei. The trapped ions architecture, as the second
example, focuses both electronic and nuclear vibrations, such as vibronic
coupling. Third, the neutral atoms architecture primarily focuses on
electronic structure and lattice models, but increasingly being explored
for nuclear motions through analog simulation of vibronic Hamiltonians.
On the other hand, variational quantum eigensolver (VQE), that is one
kind of VQAs, is useful in solving eigenvalue problem, such as electronic
structure \cite{cao19:10856}, as illustrated by Figure \ref{fig:vqe-for-quan-chem}.
Moreover, Tavernelli and co-workers
\cite{oll20:260511,oll20:043140,oll21:4229} developed quantum computational
algorithms, aiming to alleviate the exponential scaling inherent to
quantum dynamics calculations. Recently, Zhao and co-workers \cite{zhu26:xxx}
validated the feasibility of quantum simulation for photo-dissociation
dynamics of NOCl on classical computer. In this work, we will explore the
implementation possibility of quantum simulation for quantum molecular
dynamics, in particular for reaction dynamics. To analyze this
aspect, we will examine (1) the conjugacy relation between quantum
simulator and the target molecular system, (2) the wave function
correspondence in quantum algorithm and classical algorithm, say
the multi-layer multi-configuration Hartree (ML-MCTDH) method designed
for multi-dimensional quantum dynamics
\cite{wan03:1289,man08:164116,ven11:044135,wan15:7951,zha25:20397},
(3) theoretical problems of real-valued classical algorithms, and
finally (4) geometric phase arisen from the separation among the
degrees of freedom (DOFs). 

As is well known, the aforementioned points (1) and (2) play fundamental
roles in quantum simulation of quantum many-body systems, and points
(3) and (4) are theoretical issues that might introduce problems in
classical and quantum computing. In this work, we will mainly focus on
points (3) and (4) by analysis of points (1) and (2). We will discuss
gauge freedom in high-dimensional quantum molecular dynamics that has
been introduced recently, and then consider possibility of quantum
simulation of molecular reaction dynamics. It is worth noting that,
although satndard quantum theory was formulated in terms of operators
acting on complex Hilbert spaces, current algorithms for both electornic
structure and quantum molecular dynamics were developed based on real
numbers. Over the past years, it was experimentally found
\cite{wei20:060406,ren21:625,wu22:140401,li22:040402} that real-valued
quantum mechanics can be falsified. In contrast, Barrios Hita and
co-workers \cite{bar26:240202} theoretically argued that real-valued
quantum mechanics cannot be falsified. Considering a fixed $N$-dimensional
basis set for understanding such contradiction, a map from the complex
representation $\vert\psi\rangle=(c_1\;c_2\;\cdots\;c_N)^{\mathrm T}$
to the real one of a pure state can be given by
$\mathcal{S}:\mathbb{C}^N\to\mathbb{R}^{2N}:
\vert\psi\rangle\mapsto\mathrm{Re}\vert\psi\rangle\otimes\vert0\rangle_F
+\mathrm{Im}\vert\psi\rangle\otimes\vert1\rangle_F=\vert\tilde{\psi}\rangle$,
where the subscript $F$ stands for flag whose two real DOFs indicate
the real and imaginary part of $\vert\psi\rangle$. Although $\mathcal{S}:
\vert\psi\rangle\mapsto\vert\tilde{\psi}\rangle$ formally resembles the
structure of an entangled state, $\vert\tilde{\psi}\rangle$ is a single
state and thus is not entangled. In general, the ansatz for chemical
dynamics is multi-layered linear expansion of a $m$-dimensional set of
products of single-particle functions (SPFs), called configurations 
\cite{zha25:20397}. If the SPF is futher expanded by the $N$-dimensional
basis set, a complex vector $\vert\psi\rangle\in\bigotimes_{i=1}^m\mathbb{C}_i^N$
then has $2mN$ real parameters. However, its real-valued counterpart
$\vert\tilde{\psi}\rangle\in\bigotimes_{i=1}^m\mathbb{R}_i^{2N}$ has
$(2N)^m$ parameters. This implies that the map $\mathcal{S}$ is not well
defined. To overcome this problem, one must note that each of independent
subsystems can have an undetectable phase. Due to linearity in each
subsystem, a global phase is equal to the product of the individual
phases and hence can be split up between the subsystems in an arbitrary
way, also denoted by phase kickback.

The geometric phase is introduced by separation of DOFs in some special
cases \cite{men22:16415,zha25:20397}. For a high-dimensional molecule,
an appropriate separation of nuclear coordnates is usually required to
simplify the dynamical model and to save computational cost
\cite{wan03:1289,man08:164116,ven11:044135,wan15:7951,men15:164310,men17:184305,men21:2702,men22:16415,zha25:20397}.
In general, a normalized state $\vert\psi\rangle$ is equilvalent to
$\vert\psi\rangle\exp(\mi\varphi)$ with the global phase $\varphi\in\mathbb{R}$.
Then, the aforementioned map $\mathcal{S}$ satisfies \cite{bar26:240202}
$\mathcal{S}:\vert\psi\rangle\exp(\mi\varphi)\mapsto
\mathrm{Re}\vert\psi\rangle\otimes(\vert0\rangle_F\cos(\varphi)+
\vert1\rangle_F\sin(\varphi))+\mathrm{Im}\vert\psi\rangle\otimes
(-\vert0\rangle_F\sin(\varphi)+\vert1\rangle_F\cos(\varphi))=
R_F(\varphi)\mathcal{S}(\vert\psi\rangle)=R_F(\varphi)
\vert\tilde{\psi}\rangle$, which means that $\varphi$ or U(1) ambiguity
in a complex state vector corresponds to an SO(2) ambiguity in the real
formulation. The physical indistinguishability of
$\vert\tilde{\psi}\rangle$ and $R_F(\varphi)\vert\tilde{\psi}\rangle$
is guaranteed if and only if the real measurement operators lead to
identical outcome probabilities.
Then, Barrios Hita and co-workers \cite{bar26:240202} found that a
pure state in real quantum mechanics can be defined as separable if
and only if it is either a product state or equivalent to a product
state. Otherwise, it is entangled. Moreover, the embedding of the
local actions on subsystems into the operators for the composite
system is well defined \cite{bar26:240202}. The reduced density
operator in the real-valued formalism is still given via the (real)
partial trace. Keeping the above results in the mind, this work focuses
on implementation possibility of quantum simulation for quantum molecular
dynamics by reviewing aforementioned theoretical problems and previously
reported studies on this aspect.

The rest of this paper is organized as follows. Section \ref{sec:theory}
describes details of the quantum simulation for chemical dynamics. Section
\ref{sec:sm-chem} gives implementations and discussions on quantum simulation
for quantum molecular dynamics. Finally, Section \ref{sec:con} concludes
with a summary. Unless otherwise specified, the atomic units are always
used. %, while uppercase and lowercase Greek letters
%represent quantities of the system and quantum simulator, respectively.

%----------------------------------------------
% Theory
%-----------------------------------------------
\section{Theory\label{sec:theory}}

%---------------------------------
\subsection{Quantum Simulation for Chemical Dynamics\label{sec:qm-quan-md}}

As is well known, unlike classical computers, the quantum computer is
an ensemble of well-defined qubits, on which quantum simulations can
be performed, called quantum computing. Quantum simulation on evolution
of the molecular system from $\vert\Psi(0)\rangle$ to $\vert\Psi(t)\rangle$
with unitary transformation $\hat{U}=\exp(-\mi\hat{H}t)$ requires the
well-designed quantum simulator that evoles from $\vert\psi(0)\rangle$
to $\vert\psi(t)\rangle$ by $\hat{u}=\exp(-\mi\hat{h}t)$. The quantum
simulator is designed such that there is a mapping between the simulator
and the system, in particular the mappings $\vert\Psi(0)
\rangle\leftrightarrow\vert\psi(0)\rangle$, $\vert\Psi(t)\rangle
\leftrightarrow\vert\psi(t)\rangle$, and $\hat{U}\leftrightarrow\hat{u}$.
While the molecular system may not be controllable, the quantum simulator
is. The simplest quantum simulation is quantum molecular dynamics based
on a simplified counterpart (that is reduced model) for the full-dimensional
system whose direct simulation would be computationally unfeasible. This
is a kind of analogue quantum simulation (AQS) \cite{ha25:19423},
corresponding to digital quantum simulator (DQS). For
a simple example, we assume that a reaction system is composed by the
reactive and non-reactive modes, associated with large-amplitude and
perodic motions during the reaction. These modes are often determined
through chemical experiences, which are also adopted for reducing
dimension of the reaction system. The reduced and full dimensional
systems are thus called in the conjugacy relation. Taking the
H + CH$_4$ $\to$ H$_2$ + CH$_3$ reaction dynamics as an example, the
CH$_3$ fragment of the CH$_4$ is supposed to belong to the C$_{3v}$ group,
reducing the system dimension of $3\times6-6=12$ to $12-3=9$. Taking the
dissociative chemsorption dynamics of H$_2$O on Ni(111) as the second
example, the lattice effects arisen from vibrations of the surface atoms are
often totally ignored to largely save computational cost because only
dimension of $3\times3=9$ is considered.

Now, we have to give the concept of the conjugacy. To do this, let us
revisit some concepts. A discrete-time dynamical system $(X,f)$ consists
of a non-empty set $X$
and a mapping $f:X\to X$, where the composition of mappings $f\circ f$
satisfies that (1) the $n$-fold composition is denoted by $f^n=f\circ\cdots\circ f$
with $f^0$ the identity map and (2) if the inverse $f^{-1}$ exists then
$f^{-n}=f^{-1}\circ\cdots\circ f^{-1}$. Thus, if $f^{-1}$ exists, then
the iterates $\{f^n\}$ form a group under composition due to $f^n\circ f^m=f^{n+m}$.
Otherwise, they form a semigroup. Next, a continuous-time dynamical system
consists of a space $X$ together with a one-parameter family of maps
$\{f^{t}:X\to X,\;t\in\mathbb{R}\;\mathrm{or}\;t\in\mathbb{R}_{0}^{+}\}$
forming a group or semigroup, such that $f^{t+s}=f^{t}\circ f^{s}$ and
$f^{0}$ is the identity. If $t\in\mathbb{R}$, the system is called a
flow. If $t\in\mathbb{R}_{0}^{+}$, it is called a semiflow. For a flow,
since $f^{-t}=(f^{t})^{-1}$, each map $f^{t}$ is invertible. For a fixed
$t_{0}$, the iterates $(f^{t_{0}})^{n}=f^{t_{0}n}$ define a discrete-time
dynamical system. For point $x\in X$, the positive semiorbit is defined
as $\mathscr{O}_{f}^{+}(x)=\bigcup_{t\ge 0}f^{t}(x)$. In the invertible
case, the negative semiorbit is $\mathscr{O}_{f}^{-}(x)=\bigcup_{t\le0}f^{t}(x)$.
The union $\mathscr{O}_{f}(x)=\mathscr{O}_{f}^{+}(x)\cup\mathscr{O}_{f}^{-}(x)
=\bigcup_{t}f^{t}(x)$ is called the orbit of $x$. Let $f^{t}:X\to X$
and $g^{t}:Y\to Y$ be two dynamical systems. A surjective map $\pi:Y\to X$
satisfying $f^{t}\circ\pi=\pi\circ g^{t}$ for all $t$ is called a
semiconjugacy from $g$ to $f$. If $\pi$ is invertible, it is called
a conjugacy. If a conjugacy exists between two dynamical systems,
they are said to be conjugate. In this context, conjugacy is an
equivalence relation. To study a complicated dynamical system,
one often seeks a simpler model system and uses the conjugacy to
analyse the motion of the original system.
Therefore, quantum simulation requires a conjugacy relation between
the molecular system and a quantum computing device of qubits, such
that measurements on the device serve as a substitute for measurements
on the original system.

Now, we turn to time-dependent VQAs, where a
time-dependent variational trial $\vert\psi[\boldsymbol{\vartheta}(t)]\rangle$
approximates the solution $\vert\Psi(t)\rangle$. To prepare the trial
$\vert\psi[\boldsymbol{\vartheta}(t)]\rangle$ on a quantum computer, a
parametrized unitary $\hat{u}(\boldsymbol{\vartheta}(t))$, that is a
quantum circuit, is required to act onto a reference qubit state
$\vert\psi(0)\rangle$, namely $\vert\psi[\boldsymbol{\vartheta}(t)]\rangle
=\hat{u}(\boldsymbol{\vartheta}(t))\vert\psi(0)\rangle$. Since the
parameters enter as qubit-gate angles, all parameters must be real.
Based on complex structure of the McLachlan variational principle,
namely the Dirac-Frenkel variational principle
$\langle\delta\psi[\boldsymbol{\vartheta}(t)]\vert\mi\partial_t-\hat{h}
\vert\psi[\boldsymbol{\vartheta}(t)]\rangle=0$, one can obtain the
equation of motion (EOM)
for paramters $\boldsymbol{\vartheta}(t)$ in the matrix form
$\boldsymbol{\mathcal{M}}\dot{\boldsymbol{\vartheta}}=\boldsymbol{\mathcal{V}}$
with the elements
$\mathcal{M}_{ij}=\mathrm{Re}(\langle\partial_i\psi\vert\partial_j\psi\rangle+
\langle\partial_i\psi\vert\psi\rangle\langle\partial_j\psi\vert\psi\rangle)$
and $\mathcal{V}_i=\mathrm{Im}(\langle\partial_i\psi\vert\hat{h}\vert\psi\rangle-
\langle\partial_i\psi\vert\psi\rangle\langle\psi\vert\hat{h}\vert\psi\rangle)$,
where $\partial_i=\partial/\partial\vartheta_i$. In principle, this EOM
for $\boldsymbol{\vartheta}(t)$ can be integrated with any numerical
solver. By quantum computing, this algorithm becomes the quantum-EOM
(qEOM) method, as schematically illustrated by Figure \ref{fig:quan-eom-chem}.
However, the stability of the dynamics simulation depends on the system
and setups, such as errors of numerical integration and quantum hardware
noise on the integration \cite{oll20:260511,oll20:043140,oll21:4229}. It is worth noting that the
above derivations are obviously valid for either classical or quantum
computer. The distinction is whether the ansatz preparation and the
matrix evaluation adopt quantum circuits or not. The EOM for $\boldsymbol{\vartheta}$
indicates a near-term approach. Other alternative near-term approaches
are the adaptive variational quantum dynamics simulation, the subspace
variational quantum simulation, and the variational fast forwarding.
Therefore, the first step of time-dependent VQAs is preparation of
quantum state based on the quantum device. The second step is time
evolution in which propagation of one SPF is simulated by one qubit.
These steps are same as those in the classical computation.

Next, considering $N$ qubits to build $\vert\psi[\boldsymbol{\vartheta}(t)]\rangle$, the
Hilbert space is given by $\mathbb{H}_A\otimes\mathbb{H}_B\otimes\cdots\otimes\mathbb{H}_N$,
where the labels $J=\{A,B,\cdots,N\}$ are physically used to demarcate
the space of each qubit. A configuration basis
$\vert I\rangle=\vert i_A\rangle\otimes\vert i_B\rangle
\otimes\cdots\otimes\vert i_N\rangle=\vert i_Ai_B\cdots i_N\rangle$ can
be thus defined by Kronecker product, where $I=\{i_A,i_B,\cdots,i_N\}$
represents a set of indices. Due to $\{i_K=0,1\vert K\in J\}$,
there are $2^N$ basis states for $N$ qubits and hence the basis set is
given by $\{\vert I\rangle\}_{i=I}^{2^N-1}$. A general state
$\rho_{A,B,C,\cdots,N}=\rho_J$ of the multi-qubit device would again
correspond to a positive semi-definite operator with unit trace
$\rho_{A,B,C,\cdots,N}=\rho_J=\sum_{I,K=0}^{2^N-1}\vert I\rangle\rho_{IK}\langle K\vert$,
where $\rho_{IJ}\in\mathbb{C}^2$ represents element of representation
matrix. One can further define a reduced state for each of the qubits
through partial tracing of the state $\rho_J$ as $\rho_{J^K}=\mathrm{Tr}_{J^K}(\rho_J)
\in\mathscr{L}(\mathbb{H}_{J^K})$, where $J^K=\{A,B,\cdots,K-1,K+1,\cdots,N\}$.
This operation for the $K$-th qubit is often called the reduced density
operator. It is completely positive trace preserving map and generates
valid state of the $K$-th qubit. If the general state $\rho_{A,B,C,\cdots,N}=\rho_J$
is pure, then the associated vector $\vert\psi\rangle_J=
\vert\psi\rangle_{A,B,C,\cdots,N}\in\mathbb{H}_A\otimes\mathbb{H}_B
\otimes\cdots\otimes\mathbb{H}_N$ in the $N$-qubit computational basis
$\{\vert I\rangle\}_{I=1}^{2^N-1}$ is given by
\begin{equation}
\big\vert\psi[C_{i_Ai_B\cdots i_N}]\big\rangle_{A,B,C,\cdots,N}
=\sum_{i_A=0}^1\sum_{i_B=0}^1\cdots\sum_{i_N=0}^1
\big\vert i_Ai_B\cdots i_N\big\rangle
C_{i_Ai_B\cdots i_N}\;\mathrm{or}\;
\big\vert\psi[C_I]\big\rangle_J=\sum_{I=0}^{2^N-1}
\big\vert I\big\rangle C_I,
\label{eq:quantum-computing-001}
\end{equation}
where $C_{i_Ai_B\cdots i_N}=C_I\in\mathbb{C}^2$ are expansion coefficients.
For a normalized state, there should exist normalization expression
$\sum_I\vert C_I\vert^2=1$.
Furthermore, if each coefficient $C_I$ is multiplicatively factorizable
into scalars characterizing the basis states of each qubit, then the
pure state $\vert\psi[C_I]\rangle_J$ given in Equation \eqref{eq:quantum-computing-001}
can be termed separately in the form
$\vert\psi[C_I]\rangle_J=\vert\phi_A\rangle\otimes\vert\phi_B\rangle\otimes\cdots\otimes
\vert\phi_N\rangle=\bigotimes_{K\in J}\vert\phi_K\rangle$, where
$\vert\phi_K\rangle\in\mathbb{H}_K$ and $J=\{A,B,\cdots,N\}$.

%-------------------------------------
\subsection{Revisit on Wavepacket Propagation on System\label{sec:ml-mctdh}}

It is worth noting that, Equation \eqref{eq:quantum-computing-001} shares
a formal similarity with the ML-MCTDH ansatz
\cite{wan03:1289,man08:164116,ven11:044135,wan15:7951,zha25:20397}.
The ML-MCTDH method has been well discussed in the literature
\cite{wan03:1289,man08:164116,ven11:044135,wan15:7951,zha25:20397}, such
that only a brief description is given as follows. By ML-MCTDH
\cite{wan03:1289,man08:164116,ven11:044135,wan15:7951,zha25:20397}, the
total time-dependent wave function is expressed in terms of a hierarchical
Tucker format, which has a tree-like structure, through a set of multi-dimensional
and time-dependent basis functions. These basis functions are themselves
expanded in an underlying multi-dimensional and time-dependent basis.
This expansional scheme is repeated until in the lowest level a
time-independent primitive basis is used, such as basis funciton of
the discrete variable representation (DVR). Thus, the tree-like
structure of an ML-MCTDH wave function is most conveniently
visualized by a plot of the aforementioned tree structure, called
ML-tree. The expansional expression of the SPFs in the $(l-1)$-th
layer is given by \cite{wan03:1289,man08:164116,ven11:044135,wan15:7951,zha25:20397}
\begin{equation}%%17
\varphi_{m}^{(\mathfrak{z}-1;\kappa_{l-1})}
\Big(Q^{(\mathfrak{z}-1;\kappa_{l-1})},t\Big)=
\sum_{j_1}^{n_1^{(\mathfrak{z})}}\cdots
\sum_{j_{p_{\kappa_l}}}^{n_{\kappa_l}^{(\mathfrak{z})}}
A_{m;j_{1},\cdots,j_{p_{\kappa_l}}}^{(\mathfrak{z})}\big(t\big)
\prod_{\kappa_l=1}^{p_{\kappa_l}}
\varphi_{j_{\kappa_l}}^{(\mathfrak{z},\kappa_l)}\Big(Q^{(\mathfrak{z};\kappa_l)}\Big)
=\sum_JA_{m;J}^{(\mathfrak{z})}\big(t\big)\cdot\Phi_J^{(\mathfrak{z})}
\Big(Q^{(\mathfrak{z};\kappa_l)}\Big),
\label{eq:ml-wf}
\end{equation}
where $\mathfrak{z}=\{l;\kappa_1,\cdots,\kappa_{l-1}\}$ and $\mathfrak{z}
-1=\{l-1;\kappa_1,\cdots,\kappa_{l-2}\}$. The symbol $l$ denotes the
layer depth and $\mathfrak{z}$ indicates a particular node in the
ML-tree structure. The logical coordinate, that is $Q^{(\mathfrak{z}-1;
\kappa_{l-1})}=\{Q_1^{(\mathfrak{z})},\cdots,Q_{p_{\kappa_l}}^{(\mathfrak{z})}\}$,
is a combination scheme of underlying coordinates and hence useful in
multi-dimensional dynamics calculations. We refer the reader to Reference
\cite{zha25:20397} for details how to design the logical coordinate set
and the ML-tree structure. Finally, it is worth noting that, if a pure
state either given in Equation \eqref{eq:quantum-computing-001} or presented
by the ML-MCTDH ansatz is not separable, such state is called as entangled
state \cite{pan12:777}. In entanglement, subsystems are described by a
single entangled state, making their individual properties indeterminate
yet perfectly correlated. The presence of entanglement is a feature wherein
computation using qubits can be different from that of the classical bit
counterparts. It is a resource that powers quantum computing.

Inserting the multi-layer wave function into the Dirac-Frenkel variational
principle, the ML-MCTDH equations of motion (EOMs) for arbitrary layering
schemes have been derived together with an algorithm for the recursive
evaluation of all intermediate quantities entering the EOM by Wang and
Thoss \cite{wan03:1289}, Manthe \cite{man08:164116}, and Vendrell and
Meyer \cite{ven11:044135}. For the top layer coefficients, the working
equations are given by \cite{wan03:1289,man08:164116,ven11:044135},
\begin{equation}%%18
\mi\frac{\partial}{\partial t}A_{1;J}^{(1)}
=\sum_I\Big\langle\Phi_J^{(1)}\Big\vert\hat{H}\Big
\vert\Phi_I^{(1)}\Big\rangle A_{1;I}^{(1)}
-\sum_{\kappa=1}^d\sum_{i=1}^{n_{\kappa}}
g_{j_{\kappa}i}^{(1;\kappa)}A_{1;j_{\kappa}}^{(1)},\quad
\Phi_{J}^{(1)}=\varphi_{j_{1}}^{(1;1)}\Big(Q_{1}^{(1)},t\Big)
\cdots\varphi_{j_{p}}^{(1;p)}\Big(Q_{p}^{(1)},t\Big),
\label{eq:eoms-ml-mctdh-000}
\end{equation}
where $g_{j_{\kappa}i}^{(1;\kappa)}$ is called by constraint in the
chemistry literature \cite{wan03:1289,man08:164116,ven11:044135} and
is a kind of K{\"a}hler metric, specifically the Fubini-Study metric.
In Equation \eqref{eq:eoms-ml-mctdh-000}, the top layer configurations
$\Phi_{J}^{(1)}$ are defined as direct products of SPFs and the
multi-index $J=j_{1},\cdots,j_{p}$ has been implicitly introduced.
The working equation for the propagation of the SPFs are formally the
same for all layers
\begin{equation}%%20
\mi\frac{\partial\varphi_{n}^{(\mathfrak{z};\kappa_{l})}}{\partial t}
=\Big(1-P^{(\mathfrak{z};{\kappa_l})}\Big)\sum_{j,m}
\Big(\rho^{(\mathfrak{z};\kappa_{l})}\Big)_{nj}^{-1}\cdot
\Big\langle\hat{H}\Big\rangle_{jm}^{(\mathfrak{z};\kappa_{l})}
\varphi_{m}^{(\mathfrak{z};\kappa_{l})},
\quad P^{(\mathfrak{z};{\kappa_l})}=
\sum_{j}\Big\vert\varphi_{j}^{(\mathfrak{z},\kappa_{l})}\Big\rangle
\Big\langle\varphi_{j}^{(\mathfrak{z},\kappa_{l})}\Big\vert,
\label{eq:eoms-ml-mctdh-001}
\end{equation}
is the projector onto the space spanned by the SPFs $\varphi_{j}^{(\mathfrak{z},\kappa_{l})}$,
where $\rho^{(\mathfrak{z},\kappa_{l})}$ is a density matrix and
$\langle\hat{H}\rangle_{jm}^{(\mathfrak{z},\kappa_{l})}$ is a matrix
element of mean-field operators acting on the SPFs $\varphi_{j}^{(\mathfrak{z},\kappa_{l})}$.
Comparing with Equation \eqref{eq:eoms-ml-mctdh-000}, the working
equation for SPFs (see Equation \eqref{eq:eoms-ml-mctdh-001}) already
adopts zero constraint which is the simplest case. Since Equations
\eqref{eq:eoms-ml-mctdh-000} and \eqref{eq:eoms-ml-mctdh-001} for
propagation are a set of coupled non-linear differential equations,
these equations can be efficiently solved using standard numerical
tools on a classical computer cluster. We refer the reader to References
\cite{wan03:1289,man08:164116,ven11:044135,wan15:7951} for details of
the ML-MCTDH algorithm in classical computing.

Next, we must consider the possibility of a conjugacy relation between
the quantum simulator and the target molecular system. Assuming that
the molecular system $f:X\to X$ evolves on space $X$
and quantum simulator $g:Y\to Y$ evolves on space $Y$, a surjective map
$\pi:Y\to X$ is a semiconjugacy from $g$ to $f$ if $f^t\circ\pi=\pi\circ g^t$.
Moreover, If the semiconjugacy $\pi$ is invertible, then it is called
a conjugacy. Since both the target system and the quantum simulator
obey the same quantum mechanical laws, the underlying dynamical systems
$f^t$ and $g^t$ are conjugate if the simulator is properly designed
making the map $\pi$ appropriate. Obviously, if only nuclear motion
is simulated, where $X$ is the configuration space, then $Y$ must be
topologically isomorphic to $Y$. Conversely, if electronic structure
is simulated, the fermionic property of electrons necessitates that
$X$ and $Y$ incorporate spin variables. In the case of simultaneous
simulation of electronic and nuclear motion, both the configuration
space (or a topologically isomorphic space) and spin variables must
be included in both $X$ and $Y$. For example, the superconducting circuits
primarily focus on electronic energies at fixed nuclei, where the
simulator is the Cooper pairs formed within the superconductor. The
trapped ions architecture, as the second example, focuses both electronic
and nuclear vibrations, such as vibronic coupling. Finally, the neutral
atoms architecture primarily focuses on electronic structure and lattice
models, but increasingly being explored for nuclear motions through analog
simulation of vibronic Hamiltonians. In the latter two devices, the
simulator states are internal electronic states of the ion or atom.
Although we have established the existence of the quantum simulator
for wavepacket propagation, theoretical problem on the relation between
two quantum subsytems remains to be addressed.
%For implementation,
%Figure \ref{fig:prop-quan-simul} illustrates an example of quantum
%circuit for the wave packet propagation based on the Marcus spin-boson
%model \cite{oll20:260511}, $\hat{H}=\hat{K}\otimes I+V_0\otimes\vert0\rangle\langle0\vert+
%V_1\otimes\vert1\rangle\langle1\vert+C\otimes\sigma_x$.

Finally, we are interested in the fact that wavepacket propagation is
a kind of unitary transformation, which constitutes a form of information
propagation. In quantum computing, the transmission and copy of the
quantum state itself must be taken into account, as it provides the
carrier through which information is conveyed. In propagating or copying
a quantum state, Diekes \cite{die82:271} demonstrated that superluminal
communication by entanglement is not possible and stressed the crucial
role of the linearity of the evolution laws in preventing causal anomalies.
To understand this point by introducing a logic contradiction, we assume
that a molecule in $\vert\Psi\rangle$ state dissociates (or called decays)
into two fragments (say H$_2$ $\to$ H + H), denoted by I and II in the
$\vert\mathrm{I}\rangle$ and $\vert\mathrm{II}\rangle$ states. Since
the molecular system is not separable, the two fragments are entangled,
that is $\vert\Psi\rangle=\vert\mathrm{I}\rangle\otimes\vert\mathrm{II}\rangle$.
Now, observer $\mathcal{A}$ has the choice to measure either property
$O_1$ or $O_2$ of fragment I. If another observer $\mathcal{B}$ observes
the state of fragment II by measurement for only $O_1$, there are then
two possibilities observed by $\mathcal{B}$ depending upon the choice
of $\mathcal{A}$. For insatnce, if $\mathcal{A}$ decided to measure
$O_1$ of I, immediately after this measurement, fragment II can be
described with an eigenstate of operator $\hat{O}_1$. This result
can be determined by subsequent observation performed by $\mathcal{B}$.
A similar result holds if $\mathcal{A}$ decided to measure $O_2$. These
two possibilities seem to provide a way for $\mathcal{B}$ to superluminally
find out which choice has been made by $\mathcal{A}$. However, taking
observation on the spin components as an example, Diekes \cite{die82:271}
demonstrated that fragment II is always in a superposition of the
eigenstates of either $\hat{O}_1$ or $\hat{O}_2$, which is independent
of the choice of $\mathcal{A}$. A subsequent measurement of $O_1$ yields
a probabilistic ensemble of the eigenvalues and corresponding eigenstates
of $\hat{O}_1$. In this way, there is no way for $\mathcal{B}$ to find
out, on the basis of measurements on II, what $\mathcal{A}$ has decided.
Therefore, quantum mechanics does not allow superluminal communication.
We will return to this point later.

%-----------------------------------------------
\subsection{Geometric Phase\label{sec:phase}}

Considering the map $\mathcal{S}:\vert\psi\rangle\mapsto\vert\tilde{\psi}\rangle$
for $d$-dimensional system whose $m$ subsystems (or called modes) are simply
supposed to be all represented by $N$-dimensional vectors
$\{\vert\psi_i\rangle\in\mathbb{C}^N\}_{i=1}^m$, one can find that the
real and complex quantum theories are isomorphic \cite{bar26:240202}.
However, a natural extension of $\mathcal{S}$ to $\mathcal{S}^{\otimes m}:
\bigotimes_{i=1}^m\mathbb{C}_i^N\to\bigotimes_{i=1}^m\mathbb{R}_i^{2N}:
\vert\psi_1\rangle\otimes\cdots\otimes\vert\psi_m\rangle\mapsto
\mathcal{S}(\vert\psi_1\rangle)\otimes\cdots\otimes\mathcal{S}(\vert\psi_m\rangle)
=\vert\tilde{\psi}_1\rangle\otimes\cdots\otimes\vert\tilde{\psi}_m\rangle$
is not well defined because a vector in $\bigotimes_{i=1}^m\mathbb{C}_i^N$
has $2mN$ real parameters, while a vector in $\bigotimes_{i=1}^m\mathbb{R}_i^{2N}$
has $(2N)^m$ parameters. To overcome this problem, we must note that each
of independent subsystems can have an undetectable phase. Due to linearity
in each subsystem, a global phase is equal to the product of the individual
phases and hence can be split up between the subsystems in an arbitrary way,
that is phase kickback. Since $\mathcal{S}$ is invertible and its extension
$(\mathcal{S}^{-1})^{\otimes m}$ is well defined, we further consider the
invertible map
\begin{equation}
\mathcal{Q}:\bigotimes_{i=1}^m\mathbb{C}_i^N\to
\bigotimes_{i=1}^m\mathbb{R}_i^{2N}/
\ker\Big[\big(\mathcal{S}^{-1}\big)^{\otimes m}\Big],
\label{eq:phase-real-qm-03}
\end{equation}
where the kernel of $(\mathcal{S}^{-1})^{\otimes m}$ is employed, as
two elements in the domain of $(\mathcal{S}^{-1})^{\otimes m}$ have
the same image. Each element in the quotient space in the right-hand
side of Equation \eqref{eq:phase-real-qm-03} is an equivalence class
$[\vert\psi\rangle]$, where $\vert\phi\rangle\sim\vert\psi\rangle$
if $\vert\psi\rangle-\vert\phi\rangle\in\ker[(\mathcal{S}^{-1})^{\otimes m}]$.

Thus, one can distinguish linearly independent equivalence classes
$[\vert r_i\rangle\otimes\vert0_1\cdots0_i\cdots0_N\rangle_F]$ and
$[\vert r_i\rangle\otimes\vert0_1\cdots1_i\cdots0_N\rangle_F]$ for
$\{\vert r_i\rangle\}_{i=1}^N$ a basis of $\mathbb{R}^d$ and can
define \cite{bar26:240202}
\begin{equation}
\big(\mathbb{R}^d\big)^{\otimes N}\otimes
\big(\mathbb{R}^2\big)^{\otimes N}/
\ker\Big[\big(\mathcal{S}^{-1}\big)^{\otimes m}\Big]=
\mathrm{span}\left(\bigcup_{i=1}^{d^N}
\left\{\Big[\vert r_i\rangle\otimes\vert0_1
\cdots0_i\cdots0_N\rangle_F\Big],
\Big[\vert r_i\rangle\otimes\vert0_1\cdots1_i
\cdots0_N\rangle_F\Big]
\right\}\right).
\label{eq:phase-real-qm-04}
\end{equation}
For any $\vert\psi\rangle\in(\mathbb{R}^d)^{\otimes N}\otimes
(\mathbb{R}^2)^{\otimes N}$, the canonical form of a representative of
the equivalence class $[\vert\psi\rangle]$ is $\hat{P}^{\perp}\vert\psi\rangle$,
where $\hat{P}^{\perp}$ is the projector onto
$\ker[(\mathcal{S}^{-1})^{\otimes m}]^{\perp}$. Then, one can have the
projection \cite{bar26:240202}
\begin{equation}
\mathcal{P}:\big(\mathbb{R}^d\big)^{\otimes N}\otimes
\big(\mathbb{R}^2\big)^{\otimes N}/
\ker\Big[\big(\mathcal{S}^{-1}\big)^{\otimes m}\Big]\to
\ker\Big[\big(\mathcal{S}^{-1}\big)^{\otimes m}\Big]^{\perp}
\subset\big(\mathbb{R}^d\big)^{\otimes m}:\big[\vert\psi\rangle\big]
\mapsto\vert\psi\rangle^{\perp}=\hat{P}^{\perp}\vert\psi\rangle.
\label{eq:phase-real-qm-05}
\end{equation}
Since $\vert\psi\rangle^{\perp}$ is the canonical representation of the
equivalence class, Equation \eqref{eq:phase-real-qm-05} means that
$\vert\psi\rangle^{\perp}$ is unique and independent of the choice of
$\vert\psi\rangle$ from the equivalence class. In this context, one can
define a pure state $\psi$ in real quantum mechanics as separable if and only
if it is either a product state or equivalent to a product state
\cite{bar26:240202}. Otherwise, it is entangled. Thus, we generally
write the most used ansatz \cite{zha25:20397} for both electronic
structure and quantum molecular dynamics in the form
\begin{equation}
\big\vert\psi[A_I]\big\rangle=\sum_{i_1=1}^{m_1}\cdots
\sum_{i_d=1}^{m_d}A_{i_1\cdots i_d}\bigotimes_{\kappa=1}^d
\big\vert i_{\kappa}\big\rangle=\sum_{i_1=1}^{m_1}\cdots
\sum_{i_d=1}^{m_d}A_{i_1\cdots i_d}\big\vert i_1\cdots
i_d\big\rangle=\sum_I^mA_I\vert I\rangle,
\label{eq:def-state-spa-geom-000}
\end{equation}
where $A_{i_1\cdots i_d}=A_I$ is expansional coefficients, while $\vert
i_{\kappa}\rangle$ means the $i_{\kappa}$-th SPF for the $\kappa$-th
degree of freedom (DOF). Moreover,
the embedding of the local actions on subsystems into the operators
for the composite system is also well defined \cite{bar26:240202}.
The reduced density operator in the real-valued formalism can be
still given via the (real) partial trace.

Equations \eqref{eq:phase-real-qm-03} and \eqref{eq:phase-real-qm-05}
indicates that real-valued quantum theory adpoted by either elctronic
structure or quantum molecular dynamics might ignore geometric phase
of subsystems. It is the separation of spacetime coordinates that
arises the phase. For instance, separation of temporal and spacial
coordinates arises the dynamical phase factor, while separation of the
DOFs leads to the geometric phase factor or called the Berry phase.
The geometric phase can be written as a path integral of the Berry
connection. For a general integral path, there always exists an
appropriate gauge transform to cancel out the geometric phase leading
to the gauge freedom  \cite{hen85:1,fro93:733,lit97:213,raw21:24154}.
To understand this point, let us assume that the total wave function
$\Psi(\mathbi{z})$ of a molecular system is
factorized into a term $\phi(\mathbi{z})$ for the subsystem mode and
a term $\chi(\mathbi{Q})$ for the bath mode, namely $\Psi(\mathbi{z})=
\phi(\mathbi{z})\chi(\mathbi{Q})$, where $\phi$ depends parametrically
on $\mathbi{Q}$ and is normalized $\langle\phi\vert\phi\rangle_{\mathbi{q}}=1$.
Deriving the EOMs of the subsystem and the bath by the variational
principle, we can obtain their effective Hamiltonians, such that
$\Psi(\mathbi{z})$ is automatically factorized during propagation. To
this end, the Hamiltonian operator is defined as a summation
$\hat{H}(\mathbi{z})=\hat{T}_{\mathrm{bath}}(\mathbi{Q})+
\hat{T}_{\mathrm{ss}}(\mathbi{q})+
V_{\mathrm{ss}}(\mathbi{q},\mathbi{Q}_0)+
V_{\mathrm{bath}}(\mathbi{Q})+V_{\mathrm{coul}}(\mathbi{z})$, where
$\hat{T}_{\mathrm{bath}}$ and $\hat{T}_{\mathrm{ss}}$ are the kinetic
energy operators (KEOs) for the bath and the subsystem, respectively,
while $V_{\mathrm{ss}}$ is the potential energy surface (PES) of the
subsystem with the bath fixed at some special point $\mathbi{Q}_0$.
In addition, $V_{\mathrm{bath}}$ is the PES for the bath, and $V_{\mathrm{coul}}$
is the coupling term.

Letting $\mu_1$ and $\mu_2$ be Lagrange parameters and noting normalization
conditions $\langle\phi\vert\phi\rangle_{\mathbi{q}}=1$ and
$\langle\Psi\vert\Psi\rangle_{\mathbi{q}}=1$, the optimization target
reads
\begin{equation}
\mathcal{J}\big[\phi,\chi\big]=\big\langle\phi\chi\big\vert
\hat{H}\big\vert\phi\chi\big\rangle_{\mathbi{q}}+\mu_1\Big(1-
\big\langle\phi\chi\big\vert\phi\chi\big\rangle_{\mathbi{q}}\Big)+
\mu_2\Big(1-\big\langle\chi\big\vert\chi\big\rangle_{\mathbi{Q}}\Big).
\label{eq:wf-factorization-003}
\end{equation}
Varying $\mathcal{J}$ with respect to $\phi$ and $\chi$ and setting
to zero, one can obtain
\begin{equation}
\mi\frac{\partial\phi}{\partial t}=\Big(\hat{H}-
\big(\boldsymbol{\nabla}_{\mathbi{Q}}\ln\chi\big)
\boldsymbol{\nabla}_{\mathbi{Q}}\Big)\phi=
\hat{H}_{\mathrm{ss}}\phi,\quad
\mi\frac{\partial\chi}{\partial t}=\Big(\hat{T}_{\mathrm{slow}}+E\Big)
\chi=\hat{H}_{\mathrm{bath}}\chi,\quad
E=\big\langle\phi\big\vert\hat{H}_{\mathrm{ss}}\big\vert\phi\big\rangle_{\mathbi{q}}.
\label{eq:wf-factorization-011}
\end{equation}
In deriving Equation \eqref{eq:wf-factorization-011}, we find $\mu_2=0$
and $\hat{H}_{\mathrm{bath}}\chi=\mu_1\chi$ for time-independent systems.
The effective Hamiltonian operator $\hat{H}_{\mathrm{ss}}$ has an additional
term $-(\boldsymbol{\nabla}_{\mathbi{Q}}\ln\chi)\boldsymbol{\nabla}_{\mathbi{Q}}$
that arises a phase in $\phi$. The gradient of the logarithm
$\boldsymbol{\nabla}_{\mathbi{Q}}\ln\chi$ acts as an emergent gauge
potential that couples directly to the momentum associated with
$\mathbi{Q}$. This potential is not externally imposed but arises
naturally from the entanglement between the $\mathbi{q}$ and
$\mathbi{Q}$ modes. To understand this point, we would like to give
comparisons of the geometric characteristics of the nuclear wave
function with those of the electronic wave function by several criteria,
as given in Table \ref{tab:geom-comp}.
Thus, motions of the subsystem evolve not in a flat background parameter
space, but in a curved parameter space shaped by motions of the bath or
seen as environmental entanglement. The curvature of such curved parameter
space $F(\mathbi{Q})$ is defined by the Berry connection
$A(\mathbi{Q})=\mi\langle\Psi\vert\partial_{\mathbi{Q}}\vert\Psi\rangle_{\mathbi{q}}$
as $F(\mathbi{Q})=\partial_{\mathbi{Q}}\times A(\mathbi{Q})$. It is
such evolution in a curved parameter space that gives rise to the
geometric phase, allowing us to find whether the direction of a vector
is changed when it parallel transports along a path in the vector space.
Quantitatively, the Chern number in the form of
$C\propto\int_{\Gamma}\mathrm{d}\mathbi{Q}F(\mathbi{Q})=
\int_{\Gamma}\mathrm{d}\mathbi{Q}(\partial_{\mathbi{Q}}\times A(\mathbi{Q}))$
counts the number of times the Berry curvature ``wraps'' around the
parameter space, providing the number of the pathological points.
%Similar to the spontaneous symmetry breaking of electronic states
%in high-symmetry molecular systems, the aforementioned geometric
%characteristics of the nuclear motions in parameter space imply a
%symmetry-based description on quantum molecular dynamics.

%-------------------------------------------
\subsection{The Berry and Wilczek-Zee Formula\label{sec:gauge-inter}}

In the Berry formula, the geometric phase is the holonomy in a Hermitian
line bundle since the adiabatic theorem naturally defines a connection
in such a bundle
\cite{aha91:818,boh92:53,boh91:1206,boh92:977,ken97:2431,ken03:6739,sim83:2167,sim88:1725,avr89:595,aga90:6924,ana92:307,mea92:51,mar22:206002,mar24:10416,mar24:243002}.
As given in Section \ref{sec:phase}, the quantum states $\Psi_i$ and
$\Psi_i\exp(\mi\gamma_i)$ in the Hilbert space $\mathbb{K}$ cannot be
distinguished, denoted by $\Psi_i\sim\Psi_i\exp(\mi\gamma)$, where
$\exp(\mi\gamma_i)\in\mathrm{U}(1)$. Then, the proper quantum state
space of the system is defined as $\mathbb{H}=\mathbb{K}/\!\sim$ by
removing all redundant relationships denoted by the above relation
symbol ``$\sim$'' from $\mathbb{K}$. Now, letting $\mathbb{H}$ be the
base manifold, a U(1)-principle bundle $P(\mathbb{H},\mathrm{U}(1))$
can be defined. The fiber $F_{\mathbi{q}}$ at each point of $\mathbb{H}$
contains the equivalent class of states
$\pi^{-1}=\{\Psi_i=\Psi_ig\vert\Psi_i\in\mathbb{K},\;g\in\mathrm{U}(1)\}$,
where $\pi$ is a projection with local expression defined using a loca
trivialization $\phi$, that is $\pi\circ\phi(\Psi_i,\Psi_ig)=\Psi_i$.
In this context, the Berry phase is produced by parallel transport of
a quantum state in $\mathbb{H}$ along a loop $\Gamma:[0,1]\to\mathbb{H}$,
where the loop $\Gamma$ is supposed to be parameterized by parameter
$t$ (may or may not be time). A section $\sigma:\mathbb{H}\to P$ is
defined as a smooth map and satisfies $\pi\circ\sigma=1_{\mathbb{H}}$.
Choosing a specific section is equivalent to locally fixing the phase
of $\Psi_i$, namely $\sigma(\Psi_i)=\Psi_i\exp(\mi\gamma_i)$. If
$\pi\circ\tilde{\Gamma}=\Gamma$, a curve in $P$ given by
$\tilde{\Gamma}:[0,1]\to P$ is a horizontal lift of $\Gamma$, where
$\tilde{\Gamma}$ may not be a closed loop even if $\Gamma$ is. The
horizontal lift $\tilde{\Gamma}$ satisfies $\tilde{\Gamma}(1)=
\tilde{\Gamma}(0)g_{\Gamma}(1)$, where $g_{\Gamma}\in\mathrm{U}(1)$ is
a transformation on the fiber. The set of $g_{\Gamma}$ forms a subgroup
of the structure group U(1), called the holonomy group or the Berry
holonomy for convenience. Assuming $\Gamma(0)=\Psi_i$ and $\Gamma(t)=\Psi_i$,
the horizontal lift $\tilde{\Gamma}(t)$ defines a section $\tilde{\Gamma}(t)=
\sigma(\Psi_i(t))=\Psi_i(t)\exp(\mi\gamma_i(t))$. By this way, we finally
obtain the Berry phase. We refer the reader to references by Aharonov,
Bohm, Kendrick, and co-workers \cite{aha91:818,boh92:53,boh91:1206,boh92:977,ken97:2431,ken03:6739},
Simon and co-workers \cite{sim83:2167,sim88:1725,avr89:595,aga90:6924},
Anandan \cite{ana92:307}, Mead \cite{mea92:51}, and Martinazzo and
Burghardt \cite{mar22:206002,mar24:10416,mar24:243002} for further
discussions on the geometric phase and its effects.

In addition, Wilczek and Zee \cite{wil84:2111} generalized the above
Berry formula for non-degenerate cases to the case of degenerate
Hamiltonians, leading to the non-Abelian geometric phase, known as
quantum holonomy. Wilczek and Zee \cite{wil84:2111} further found
that characteristics of the gauge fields are related to energy
splittings, which may be observable in real systems. The Wilczek-Zee
formula corresponds to replacing the aforementioned Abelian $\mathrm{U}(1)$-principal
bundle with a non-Abelian $\mathrm{U}(N)$-principal bundle whose
structure group $\mathrm{U}(N)$ acts on the $N$-fold degenerate eigenspace of
the system Hamiltonian. In this case, the relevant state space is no
longer $\mathbb{H}=\mathbb{K}/\mathrm{U}(1)$ as in the Berry formula,
but rather the Grassmannian manifold $\mathrm{Gr}(N,\mathbb{K})$
parameterized by parameter $t$ (may or may not time). Letting
$\mathrm{Gr}(N,\mathbb{K})$ be base manifold, each point in this base
manifold corresponds to an $N$-dimensional degenerate eigenspace of
the system Hamiltonian. Over $\mathrm{Gr}(N,\mathbb{K})$, we can define
a $\mathrm{U}(N)$-principal bundle, where the fiber over a point consists
of all orthonormal bases of that degenerate subspace. A section of this
bundle corresponds to a local choice of basis for the degenerate eigenspace,
and a connection on this bundle is defined by the matrix-valued Berry
connection $A_{ij}=\langle\Psi_i\vert\partial_{\mathbi{Q}}\vert\Psi_j\rangle$,
with $\{\Psi_i\}_{i=1}^N$ being an orthonormal basis of the degenerate
subspace.
%Under a local gauge transformation
%$g\in\mathrm{U}(N)$, the connection transforms as
%$\mathcal{A}\to g^{-1}\mathcal{A}g+g^{-1}dg$, which is 
%precisely the transformation law of a non-Abelian gauge potential.
In this non-Abelian framework, the quantum holonomy is the parallel
transport operator obtained by integrating this connection along a
closed loop $\Gamma$ in the base manifold $\mathrm{Gr}(N,\mathbb{K})$.
All holonomies generated by the connection $\mathbi{A}$ form a subgroup
of $\mathrm{U}(N)$, known as the holonomy group $\mathrm{Hol}(\mathbi{A})$.

Specifically, for a loop $\Gamma:[0,1]\to\mathrm{Gr}(N,\mathbb{K})$
with $\Gamma(0)=\Gamma(1)$, where $\Gamma$ is again supposed to be
parameterized by parameter $t$, the quantum holonomy is the loop
integral
\begin{equation}
\mathbi{h}_{\Gamma}=\mathcal{P}\exp\left(-\oint_\Gamma\mathbi{A}
\mathrm{d}\mathbi{Q}\right)\in\mathrm{U}(N),
\label{eq:loop-int-wil-zee}
\end{equation}
where $\mathcal{P}$ denotes path-ordering. Equation \eqref{eq:loop-int-wil-zee}
indicates that the quantum holonomy is a non-Abelian unitary matrix
that mixes the states within the degenerate subspace. The horizontal
lift of the loop $\Gamma$ in the principal $\mathrm{U}(N)$-bundle is
a curve $\tilde{\Gamma}$ that starts at some initial basis and ends at
a basis related by the holonomy matrix, namely $\tilde{\Gamma}(1)=
\tilde{\Gamma}(0)\cdot\mathbi{h}_{\Gamma}$. The holonomy is the geometric
manifestation of the non-Abelian gauge field, called the Wilczek-Zee
connection, whose curvature is given by the field strength which
generalizes the Berry curvature from an Abelian 2-form to a matrix-valued
2-form. The Ambrose-Singer theorem then ensures that the Lie algebra of
the holonomy group $\mathrm{Hol}(\mathbi{A})=\mathrm{U}(N)$ is generated
by these curvature components, linking the local geometry of the parameter
space to the global and geometric unitary transformations on the quantum
state space.
Since both geometric phases and quantum holonomies are global quantities
depending only on the evolution paths of systems, quantum gates based
on them possess built-in resilience to certain kinds of errors \cite{fuj01:75}.
In other words, both Abelian and non-Abelian holonomies provide the
mathematical foundation for geometric and holonomy quantum computation,
where quantum gates are realized as holonomies of a gauge connection
\cite{fuj01:75}, offering intrinsic robustness against control errors.
Kwek, Sj{\"o}qvist, Tong, and co-workers \cite{zha23:1} represented an
overview of the theoretical and experimental progress for constructing
geometric and holonomic quantum gates and how to combine them with other
error-resistant techniques.

In addition to aforementioned geometric theory for Abelian and non-Abelian
holonomies, the practical description of chemical dynamics frequently
requires a theoretical framework that interpolates between these two
cases. In details, the quantum states in the relevant state space are
either non-degenerate or only approximately so, while diabatic
transformations are allowed among them. This requires a formalism that
consistently incorporates both the Abelian and non-Abelian holonomies.
According to the derivation in the end of Section \ref{sec:phase}, a
new mechanism of inter-/intra-molecular vibrational energy redistribution
(IVR) can be found for bi-/uni-molecular processes. This might be helpful
to deeply understand reaction dynamics with multi-rovibrational states.
To this end, we extend Equation \eqref{eq:wf-factorization-011} to the
time-dependent equations in matrix form,
\begin{equation}
\mi\frac{\partial}{\partial t}\boldsymbol{\Psi}(t)=\left[
-\frac{1}{2}\Big(\boldsymbol{\nabla}_{\mathbi{Q}}
+\mathbi{A}\Big)^2+\mathbi{E}\right]\boldsymbol{\Psi}(t)=
\mathbi{H}_{\mathrm{adia}}\boldsymbol{\Psi}(t),\quad
\mi\frac{\partial}{\partial t}\tilde{\boldsymbol{\Psi}}(t)=
\left(-\frac{1}{2}\nabla_{\mathbi{Q}}^2+\mathbi{V}\right)
\tilde{\boldsymbol{\Psi}}(t)=\mathbi{H}_{\mathrm{dia}}
\tilde{\boldsymbol{\Psi}}(t),
\label{eq:new-adi-dia-rep-016}
\end{equation}
In Equation \eqref{eq:new-adi-dia-rep-016}, $\mathbi{H}_{\mathrm{adia}}$
and $\mathbi{H}_{\mathrm{dia}}$ are time-independent Hamiltonian matrices
in the adiabatic and diabatic representations, respectively, in which
the states in the relevant state space are represented by $\boldsymbol{\Psi}$
and $\tilde{\boldsymbol{\Psi}}$. However, the relation between $\mathbi{E}$
and $\mathbi{V}$ is usually unknown for high-dimensional systems. To
overcome this problem, one can adopt an appropriate nonadiabatic model
leading to
\begin{equation}
\mi\frac{\partial}{\partial t}\tilde{\boldsymbol{\Psi}}(t)=
\left(-\frac{1}{2}\nabla^2_{\mathbi{Q}}+\mathbi{U}\right)
\tilde{\boldsymbol{\Psi}}(t)=\mathbi{H}_{\mathrm{nonadia}}
\tilde{\boldsymbol{\Psi}}(t),
\label{eq:new-adi-dia-rep-017}
\end{equation}
where $\mathbi{H}_{\mathrm{nonadia}}$ and $\mathbi{U}$ mean nonadiabatic
Hamiltonian and interaction matrices, respectively. Thus, the IVR
process requires the aid of motions of the slow mode $\mathbi{Q}$
because it is arisen from the couplings between two separated modes.
A definitive measure of the IVR process could be provided by the time
evolutions of a limited number of accessible mode-specific evolutions.
However, it is often too simplistic to fully understand IMR taking into
account all the couplings. The IVR mechanism through Equations \eqref{eq:new-adi-dia-rep-016}
and \eqref{eq:new-adi-dia-rep-017} can well overcome this problem.

%-----------------------------------------------------
% Results and Discussions
%---------------------------------------------------
\section{Implementations and Discussions\label{sec:sm-chem}}

%-------------------------------
\subsection{Hamiltonian in the Suzuki-Trotter Form\label{sec:hamiltonian-formu}}

As is well known, the Hamiltonian operator is prerequisite of quantum
dynamics. In classical computing, $d$-dimensional wavepacket propagation
usually requires the Hamiltonian in either sum-of-products (SOP) form
or canonical polyadic (CP) decomposition \cite{zha25:20397},
\begin{align}
\hat{H}^{(\mathrm{SOP})}&=\hat{T}+V^{(\mathrm{SOP})}=
-\frac{1}{2\sqrt{\vert\tilde{\gamma}\vert}}
\sum_{\kappa,\kappa'=1}^d\frac{\partial}{\partial q^{(\kappa)}}
\left(\sqrt{\vert\tilde{\gamma}\vert}\tilde{\gamma}^{\kappa\kappa'}
\frac{\partial}{\partial q^{(\kappa')}}\right)+
\sum_{j_1=1}^{m_1}\sum_{j_2=1}^{m_2}\cdots\sum_{j_d=1}^{m_d}
A_{j_1j_2\cdots j_d}\prod_{\kappa=1}^dv^{(\kappa)}_{j_\kappa}
\Big(q^{(\kappa)}\Big),
\label{eq:ham-sop-cp-00} \allowdisplaybreaks[4] \\
\hat{H}^{(\mathrm{CP})}&=\hat{T}+V^{(\mathrm{CP})}=
-\frac{1}{2\sqrt{\vert\tilde{\gamma}\vert}}
\sum_{\kappa,\kappa'=1}^d\frac{\partial}{\partial q^{(\kappa)}}
\left(\sqrt{\vert\tilde{\gamma}\vert}\tilde{\gamma}^{\kappa\kappa'}
\frac{\partial}{\partial q^{(\kappa')}}\right)+
\sum_{r=1}^{R}A_r\prod_{\kappa=1}^dv^{(\kappa)}_r\Big(q^{(\kappa)}\Big),
\label{eq:ham-sop-cp-01}
\end{align}
where $\{q^{(\kappa)}\}_{\kappa=1}^d$ is coordinate set. In Equations
\eqref{eq:ham-sop-cp-00} and \eqref{eq:ham-sop-cp-01}, function
$v^{(\kappa)}_r(q^{(\kappa)})$ or $v^{(\kappa)}_{j_\kappa}(q^{(\kappa)})$
is called single-particle potential (SPP) for the $\kappa$-th mode and
the $r$-th or $j_\kappa$ site, while $R$ is rank of the CP decomposition.
We refer readers to Reference \cite{zha25:20397} for review of construction
techniques of the Hamiltonian operator in the SOP form or CP decomposition.
In quantum computing, on the other hand, Suzuki-Trotter decomposition
is the most common and straightforward approach to write the evolution
operator of quantum simulator as a sequence of simple gates. It is part
of a broader family of product formulas. Suzuki-Trotter decomposition
schemes, or Trotterizations, or splitting operator methods, are hence
approximations to operator exponentials.

For the simplest splitting of the first order, assuming a real Lie
algebra $\varg$ of infinite or finite dimension $k$ with basis
$\{\hat{h}_{\kappa}\}_{\kappa=1}^k$, the unitary evolution operator
$\hat{u}=\exp(-\mi\hat{h}t)=\exp(\mi\hat{x})$ with $\hat{x}=\sum_{\kappa=1}^l\hat{h}_{\kappa}$
is approximated by a unitary operator in the decomposition
\begin{equation}
\hat{w}=\prod_{j=1}^n\exp\Big(t_j\hat{h}_{k_j}\Big),\quad
t_j\in\mathbb{R},\quad k_j\in[l]=\{1,2,\cdots,l\},
\label{eq:bounded-condi-008}
\end{equation}
where $l\leq k$ and $l<\infty$. Since both $\hat{u}$ and $\hat{w}$ are
unitary, the basis $\{\hat{h}_{\kappa}\}_{\kappa=1}^k$ is skew-Hermitian
set and satisfies Lie bracket $[\hat{h}_{\kappa},\hat{h}_{\kappa'}]=
\hat{h}_{\kappa}\hat{h}_{\kappa'}-\hat{h}_{\kappa'}\hat{h}_{\kappa}=
\sum_{\kappa''=1}^k\gamma_{\kappa''}^{(\kappa\kappa')}\hat{h}_{\kappa''}$,
where constants $\gamma_{\kappa''}^{(\kappa\kappa')}\in\mathbb{R}$ are
the structure factors of $\varg$. Somma \cite{som16:062202} presented
an improved bound on the number of terms in the decomposition of $\hat{w}$
by exploiting the structure of the Lie algebra. Letting $\epsilon>0$
be a precision parameter, $p$ and $r$ positive integers that satisfy
\begin{align}
2N^2_p\sum_{j=2p}^{\infty}\left(\frac{f_jN_p}{r}\right)^jt^{j+1}
\beta_{j+1}\leq\epsilon,\quad
N_p&=2l5^{p-1},\quad\beta_j=\max\left\Vert
\Big[\hat{h}_{k_1},\Big[\cdots,\hat{h}_{k_j}\Big]\cdots\Big]
\hat{u}(\lambda)\big\vert\psi(0)\Big\rangle\right\Vert,
\allowdisplaybreaks[4] \nonumber \\
f_j&=\left\{\begin{array}{ll}
2, & j<N_p \\
\dfrac{6N_p}{j}, & j\geq N_p
\end{array}\right.,\quad\lambda\in\big[0,t\big],
\label{eq:bounded-condi-009}
\end{align}
there exists a unitary $\hat{w}$ as in Equation \eqref{eq:bounded-condi-008}
that satisfies
\begin{equation}
\left\Vert\Big(\hat{u}(t)-\hat{w}\Big)
\big\vert\psi(0)\Big\rangle\right\Vert\leq\epsilon,
\quad\hat{w}=\prod_{j=1}^n\exp\Big(t_j\hat{h}_{k_j}\Big),\quad n=rN_p.
\label{eq:bounded-condi-010}
\end{equation}
In general, the time evolution $\hat{u}=\exp(-\mi\hat{h}t)=\exp(\mi\hat{x})$
is split into time steps $h$. The Hamiltonian operator consists of operators
$\sum_{j=1}^n\hat{h}_{k_j}$ and the exponential of an operator over a
sub-step $a_jh$ is called a stage $\exp(a_jh\hat{h}_{k_j})$. A stage
can be interpreted as a gate if the operator $\hat{h}_{k_j}$ is local.
Together all stages makes a ramp in forward or backward direction.
However, comparing the SOP form and CP decomposition in Equations
\eqref{eq:ham-sop-cp-00} and \eqref{eq:ham-sop-cp-01} with Equation
\eqref{eq:bounded-condi-008}, the Suzuki-Trotter decomposition is only
a special and simple case. The Hamiltonian of a molecular system often
consists out of more than three non-commuting operators, while the
potential term is typically non-analytic and not automatically available
to either the SOP form or the CP decomposition. Thus, it is necessary
to split the Hamiltonian into $\mathcal{O}(l)$ local contributions
\cite{hat05:37}. This is not only beneficial for numerical simulations
by classical computing, but also essential for applying quantum gates
to quantum molecular dynamics.

In addition, the second-order Suzuki-Trotter decomposition approximates
the time evolution operator by symmetrically arranging the exponentials.
This cancels out lower-order error terms, thereby improving the overall
accuracy from first order to second order in $h/t$ (see Equation
\eqref{eq:bounded-condi-008}). For the simplest case, letting
$\hat{h}=\hat{a}+\hat{b}$ the time evolution operator is decomposed by
\begin{equation}
\hat{u}=\exp\Big(-\mi\hat{h}t\Big)=
\exp\Big(-\mi\big(\hat{a}+\hat{b}\big)t\Big)\approx
\left[\exp\left(-\frac{\mi}{2}\hat{a}h\right)
\exp\Big(-\mi\hat{b}h\Big)\exp\left(-\frac{\mi}{2}\hat{a}h\right)
\right]^{t/h}.
\label{eq:2-or-st-decomp-0}
\end{equation}
To understand the decomposition in Equation \eqref{eq:2-or-st-decomp-0},
we use the series expansion of the exponential operator and assume
small time step $h$. For sufficiently small $h$, up to the first-order
terms in $h$ we have
\begin{equation}
\exp\left(-\frac{\mi}{2}\hat{a}h\right)
\exp\Big(-\mi\hat{b}h\Big)\approx
\left(I-\frac{\mi}{2}\hat{a}h\right)
\Big(I-\mi\hat{b}h\Big)=
I-\frac{\mi}{2}\hat{a}h-\mi\hat{b}h+\mathcal{O}\big(h^2\big)
\approx I-\mi\left(\frac{1}{2}\hat{a}+\hat{b}\right)h.
\label{eq:2-or-st-decomp-1}
\end{equation}
Next, with Equation \eqref{eq:2-or-st-decomp-1} one can obtain
\begin{equation}
\exp\left(-\frac{\mi}{2}\hat{a}h\right)\exp\Big(-\mi\hat{b}h\Big)
\exp\left(-\frac{\mi}{2}\hat{a}h\right)\approx
\left[I-\mi\left(\frac{1}{2}\hat{a}+\hat{b}\right)h\right]
\left(I-\frac{\mi}{2}\hat{a}h\right)=
I-\mi\big(\hat{a}+\hat{b}\big)h+\mathcal{O}\big(h^2\big).
\label{eq:2-or-st-decomp-2}
\end{equation}
This combined product in Equation \eqref{eq:2-or-st-decomp-2} is the
single step approximation form a time step $h$, denoted by $S_2(h)$.
Up to the first order in $h$, we have limit expression
\begin{equation}
\hat{u}=\exp\Big(-\mi\hat{h}t\Big)=
\exp\Big(-\mi\big(\hat{a}+\hat{b}\big)t\Big)=
\lim_{t/h\to\infty}
\left[\exp\left(-\frac{\mi}{2}\hat{a}h\right)
\exp\Big(-\mi\hat{b}h\Big)
\exp\left(-\frac{\mi}{2}\hat{a}h\right)\right]^{t/h}.
\label{eq:2-or-st-decomp-3}
\end{equation}
Because the exponentials are arranged from $\hat{a}/2$ to $\hat{b}$ to
$\hat{a}/2$ in each small step, lower-order commutator errors cancel
out. Thus, the second-order decomposition exhibits an error that scales
like $h^2/t^2$, outperforming the first-order decomposition’s behavior
of $h/t$, making second-order splitting a popular choice in numerical
simulations for dynamics.

Let us now turn to the implementation level. Depending on the system,
the operators $\hat{a}$ and $\hat{b}$ in Equation \eqref{eq:2-or-st-decomp-3}
may represent, say the KEO and PES \cite{zhu26:xxx}, or alternatively,
the Hamiltonians of two distinct parts \cite{oll20:043140,oll20:260511,oll21:4229}.
In either case, for a typical gas-phase
reaction with $5$ to $7$ atoms (corresponding to a dimensionality of
$9$ to $15$) the Hamiltonian generally consists of on the order of
$10^3$ terms. In classical computing, such a decomposed Hamiltonian in
Equation \eqref{eq:ham-sop-cp-00} or \eqref{eq:ham-sop-cp-01} still
supports multi-dimensional wavepacket propagation, even quantum scattering,
at a reasonable computational cost \cite{zha25:20397}. However, in quantum
computing, each term in decomposed Hamiltonian must be implemented by
$\sim t/h$ quantum gates. This implies that constructing the full
Hamiltonian would require thousands of or even millions of quantum
gates, a demand that far exceeds the capabilities of current quantum
hardware and poses a serious practical bottleneck for near-term quantum
simulations. However, it is not an insurmountable barrier. Several
efforts are currently paid to overcome this bottleneck, including the
development of more gate-efficient algorithms, compilation techniques
that exploit hardware topology, and hybrid quantum-classical workflows
that reduce circuit depth by offloading part of the complexity to
classical pre-processing.

%-----------------------------
\subsection{Quantum Gates and Quantum Circuits\label{sec:quan-gate-cir}}

In classical computing, one operates classical bits using classical
logic gates, like AND, NOT, OR, and many more. Combining these classical
logic gates on a set of multiple bits creates circuits that form the
base of an operation that can be executed on a central processing unit
(CPU). Similarly, quantum gates and quantum circuits constitute the
language of quantum computing. Quantum gates are elementary operations
applied to qubits, and quantum circuits are the compositional framework
in which quantum gates are arranged to implement algorithms. A single
qubit is a two-level quantum system whose state is a unit vector in the
Hilbert space. A register of $N$ qubits lives in the tensor-product
space, whose dimension grows exponentially with $N$, as given by
Equation \eqref{eq:quantum-computing-001}. This exponential capacity
is the resource quantum algorithms exploit, and quantum gates are the
means by which it is accessed and processed coherently. Every quantum
gate is a unitary operator, reflecting the requirements that isolated
quantum evolution preserves probability and that quantum gate is reversible.
The elementary single-qubit quantum gates include the Pauli gates, the
Hadamard gate, phase gates, identity gate, and so on. To undersand
implementation for quantum molecular dynamics, in this section it is
necessary to simply revisit quantum gates and quantum circuits.

By the Hadamard gate, denoted by H-gate, one can manipulate the state
of a single qubit putting it in a superposition of $\vert0\rangle$ and
$\vert1\rangle$. By the basis set $\{(1\;0)^{\mathrm{T}},(0\;1)^{\mathrm{T}}\}$,
the H-gate is represented by unitary matrix and acts on the bases leading
to $\vert\pm\rangle$. Another kind of fundamental quantum gate is the
Pauli gates around the $x$, $y$, and $z$ axes, denoted by X-, Y-, and
Z-gates, respectively. The X-gate, also called the bitflip gate, flips
$\vert0\rangle$ to $\vert1\rangle$ and vice versa, and hence it is
similar to the classical NOT gate. Applying the X-gate to superposition
state flips the coefficients of the bases, thereby leaving $\vert\pm\rangle$
unchanged up to a global phase of $\pi$. Like the X-gate, the Y-gate
rotates the state with an angle of $\pi$ around the $y$ axis and hence
maps $\vert0\rangle$ to $\vert1\rangle\mi$ and $\vert1\rangle$ to
$-\vert0\rangle\mi$. Applying the Y-gate a superposition state maps
$\vert+\rangle$ to $\vert-\rangle$ and vice versa. The Z-gate rotates
the state with $\pi$ around the $z$ axis. Applying the Z-gate maps
$\vert0\rangle$ to $\vert0\rangle$ and $\vert1\rangle$ to $-\vert1\rangle$,
while the Z-gate flips $\vert\pm\rangle$ to $\vert\mp\rangle$. Unlike
the above gates for unitary transformations, the measurement gate, or
called M-gate, collapses the state to one of its base and is irreversible.
This implies that the M-gate is not an actual quantum gate. The M-gate
bridges the gap between the quantum world and the classical word. There
are quantum gates composed by multiple input and output qubits. Due to
reversible requirement, the number of input and output qubits are always
equal. One of widely used two-qubit gates is the controlled NOT gate
(CNOT-gate), namely controlled X-gate. The CNOT gate acts on a control
qubit and a target qubit. If the control qubit is in $\vert1\rangle$,
the target qubit is then flipped. If the control qubit is in $\vert0\rangle$,
the target qubit is left unchanged. Combining one or muliple qubits and
any number of quantum gates applied to these qubits, one can construct
the quantum circuits that are the basic blocks for a quantum algrorithm.
In general, quantum circuits are illustrated through the wires and boxes,
such as those shown in Fiugres \ref{fig:quan-eom-chem} and \ref{fig:prop-quan-simul}.
A quantum circuit can results a quantum state or a classical state. It
is worth noting that, if we need to keep quantum state at the end of
quantum circuit, then measurements should not be executed until the
calculation has fully complete. It is worth noting that, as mentioned
in Section \ref{sec:ml-mctdh}, combined qubits might be entangled and
cannot be extracted. Measuring one of these qubits, this qubit collapses
to either $\vert0\rangle$ or $\vert1\rangle$ and other entangled qubits
will collapse to the same state.

A typical quantum algorithm proceeds through three distinct stages.
First, an initial quantum state is prepared based on instructions
from a classical computer. Second, a sequence of quantum gates and
circuits, the core of the algorithm, is applied to this state using
quantum hardware. Finally, measurements are performed on the resulting
state, collapsing it into a classical bitstring that provides the output
of the computation. For instance, the VQE and Shor algorithms are two
different kinds of algorithms in quantum computing. The VQE algorithm
(see also Figure \ref{fig:vqe-for-quan-chem}), a hybrid quantum-classical
algorithm, was designed to estimate the ground-state energy of bounded
system, making it a powerful tool for electronic structure. As shown
in Figure \ref{fig:vqe-for-quan-chem}, the VQE works by preparing a
parameterized ansatz as given in Equation \eqref{eq:quantum-computing-001}
on a quantum processor, measuring the expectation value of the Hamiltonian,
and feeding the result to a classical optimizer that iteratively updates
the parameters until convergence is achieved. It requires only
shallow-to-moderate quantum circuits and is well-suited for near-term
devices, even though its performance is limited by noise, ansatz, and
the optimization landscape. On the other hand, the Shor algorithm is a
purely quantum algorithm that performs integer factorization in polynomial
time. It relies on quantum phase estimation and modular exponentiation
to find the period of a modular exponential function, which requires
deep quantum circuits, thousands of logical qubits, and fault-tolerant
error correction-resources beyond current capabilities of quantum
computing devices. In this context, VQE seeks practical utility
with current hardware by trading fidelity for feasibility.
%, the Shor
%algorithm defines the long-term goal of scalable, error-corrected
%quantum computation.

Finally, quantum teleportation is required in designing quantum
computing devices because one may need to move the quantum state from
one qubit to another qubit. This quantum state is not just $\vert0\rangle$
or $\vert1\rangle$ but includes every kind of superposition state. As
mentioned in the end of Section \ref{sec:ml-mctdh}, the non-cloning
theorem \cite{die82:271} states that it is impossible to
create an identical copy of a unknown quantum state, meaning that the
only way to copy a quantum state from one qubit to another is to repeat
every operation performed for creating it on the other qubits. Quantum
teleportation relies heavily on a shared entangled pair of qubits held
by two terminations, denoted by $\mathcal{A}$ and $\mathcal{B}$. To
teleport an arbitrary quantum state $\vert\psi\rangle$ from $\mathcal{A}$
to $\mathcal{B}$, $\mathcal{A}$ first interacts its half of the entangled
pair with the message qubit via a CNOT-gate, then applies a H-gate to
the message qubit. Then, $\mathcal{A}$ measures both qubits, yielding
one of possible classical outcomes. These classical bits are sent to
$\mathcal{B}$ via a classical computing technique. Based the recieved
classical bits, $\mathcal{B}$ applies a corresponding Pauli correction,
that is I-, X-, Z- or XZ-gate to its qubit. This operation restores the
qubit of $\mathcal{B}$ to exactly the original state $\vert\psi\rangle$.
while the original qubit of $\mathcal{A}$ is destroyed. In this way,
the quantum state of the message qubit, which may in a state of superposition,
can be teleported by only passing the classical bits of information.
If we extend this circuit to support multiple qubits carrying messages,
a lot more quantum state information can be teleported in the same way.

%----------------------------
\subsection{Implementation for Quantum Molecular Dynamics\label{sec:implem-appli-qc}}

\begin{figure}[h!]
  \centering
   \includegraphics[width=18cm]{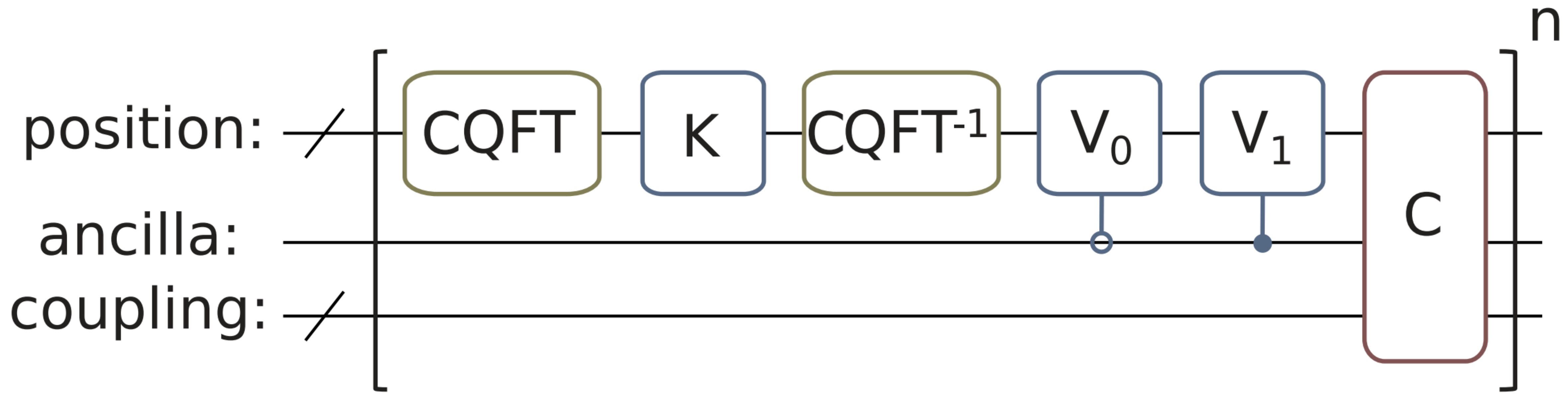}
    \caption{\figfoot}
     \label{fig:prop-quan-simul}
      \end{figure}

In this section, let us give three examples of quantum simulation on
quantum molecular dynamics. In quantum molecular dynamics, quantum
computing is currently focused on simulating the systems coupled with
electrons \cite{oll20:043140,oll20:260511,oll21:4229,zhu26:xxx}, such as non-adiabatic dynamics, photoelectron
spectrum, vibronic dynamics, and so on. This is not only because of the
intrinsic importance of these systems, but also because of practical
considerations. First, near-term quantum hardware is best suited for
problems formulated in terms of electronic Hamiltonians (see Fiugres
\ref{fig:quan-eom-chem} and \ref{fig:prop-quan-simul}). The current
generation of noisy intermediate-scale quantum (NISQ) devices with
hundreds of qubits and limited coherence
times is most efficient at simulating naturally fermionic Hamiltonians
and can be easily mapped to qubits. This is well-understood, efficient,
and accessible with current hardware. Second, the gate counts and circuit
depths required for electronic motions remain manageable with current
and near-term devices. Current electronic structure VQE algorithms (see
Figures \ref{fig:vqe-for-quan-chem} and \ref{fig:quan-eom-chem})
require circuits that scale polynomially with the number of molecular
orbitals. These circuits can be executed on current NISQ devices with
acceptable fidelity. In contrast, quantum simulating tunneling or zero-point
energy effects in reaction dynamics often requires deeper circuits and more
qubits, which are still beyond reach. Third, quantum simulators are naturally
adapted to the description of electronic structure. For instance, analog
quantum simulators were designed to mimic Hamiltonians of condensed matter
systems, which are fundamentally electronic systems. These simulators can
directly reproduce electronic-structure phenomena like superconductivity,
magnetism, and topological phases. The above considerations make electronic-structure
problems or dynamics coupled with electronic motions the most promising
entry point for quantum computing in chemistry. To see this point, we
give typical examples of quantum simulations for the molecular systems
in this section.

The first example is the non-adiabatic propagation dynamics with the
Hamiltonian of the Marcus spin-boson model \cite{oll20:260511}, that
is
\begin{equation}
\hat{H}=\hat{K}\otimes I+
V_0\otimes\big\vert0\big\rangle\big\langle0\big\vert+
V_1\otimes\big\vert1\big\rangle\big\langle1\big\vert+
C\otimes\sigma_x.
\label{eq:marcus-model-ham-01}
\end{equation}
Quantum circuit of the Hamiltonian in Equation \eqref{eq:marcus-model-ham-01}
is represented by Figure \ref{fig:prop-quan-simul}. In this Hamiltonian,
$\hat{K}$ is the KEO, $V_0$ and $V_1$ are the PESs of the first and
second diabatic states (denoted by $\vert0\rangle$ and $\vert1\rangle$),
respectively, $C$ is the coupling operator, and $\sigma_x$ is the $x$
component of the Pauli matrix. This model provides a simple description
of the electron transform. The initial state was obtained by a VQE
calculation for vibrational ground state of the $\vert1\rangle$ state.
At each iteration of the VQE, the total energy of the trial wave
function was computed by sampling both in the position and in the
momentum basis. For the latter, a centered quantum Fourier transform
(cQFT) \cite{oll20:260511} must be applied before the measurement (see
Figure \ref{fig:prop-quan-simul}). Once initialized, the initial state
is propagated by applying the Trotterized time evolution operator. The
KEO is applied in the momentum space, after performing a cQFT \cite{oll20:260511}
to transition to the momentum basis. While the quantum circuit for the
KEO can be directly applied to the first $N$ qubits, the PES must be
controlled by the state of the ancilla qubit, such that the system
evolves on $V_0$ or $V_1$ when the ancilla qubit is in the state
$\vert0\rangle$ or $\vert1\rangle$, respectively. Figure \ref{fig:marcus-quancomp}
illustartes quantum simulation results on Equation \eqref{eq:marcus-model-ham-01},
including the time-dependent population fraction $P_0$ in the product
well, linear fitting of the first ten steps of $P_0$ to approximate the
rate constants, (c) rate constants $k$ as function of the offset. In
subfigure (a), results obtained with the algorithm \cite{oll20:260511}
in classical simulations are given by dots, while the exact evolutions
with the reference coupling and the approximate coupling are given by
solid lines and dashed lines, respectively. In subfigures (a) and (b),
colors represent different offset values between $V_0$ and $V_1$. In
subfigure (c), results obtained with the algorithm \cite{oll20:260511}
are given by dots, while the exact evolution with the reference coupling
is given by crosses. The Marcus rates are shown as a dashed line for a
qualitative comparison. Therefore, the above framework proposed by
Tavernelli and co-workers \cite{oll20:260511} will pave the way toward
a understanding of femtochemistry processes via quantum computing.

The second example is the photo-dissociation dynamics of the NOCl molecule
\cite{zhu26:xxx}, whose Hamiltonian and wave function are both already
given in the SOP form. Zhao and co-workers \cite{zhu26:xxx} developed
and validated a quantum computing algorithm for such molecular dynamics
calculation through the Qiskit package \cite{jav24:2405,pat25:2508}
launched on classical computer. By integrating the split-operator
propagation method with the dilation scheme, Zhao and co-workers
\cite{zhu26:xxx} mapped the time evolution of the wave function in
a finite coordinate range onto a quantum circuit with a finite number
of qubits and extracted the autocorrelation function through the
Hadamard test (see Figure \ref{fig:nocl-quan-circuit}). Based on
numerical simulations on classical computer, the proposed quantum
algorithm accurately captures the autocorrelation function of the
NOCl system. Figure \ref{fig:nocl-spectrum}(a) illustrates validation
of this quantum simulation on classical computer against classical
benchmark calculations and confirms the good performance of the
quantum algorithm if a sufficient number of sampling shots are
performed. Furthermore, Zhao and co-workers \cite{zhu26:xxx} reported
the statistical and noise model simulation and found that insufficient
sampling shots and device noise can introduce significant fluctuations
into autocorrelation function. As shown in Figure \ref{fig:nocl-spectrum}(b),
however, the subsequent Fourier transform exerts a profound smoothing
effect. These results \cite{zhu26:xxx} demonstrated the potential
application of existing quantum devices to photo-dissociation
dynamics. Furthermore, the dilation scheme for the complex absorbing
potential involves a practical measurement trade-off.

%---------------------------------------------------------------------------
\subsection{Advantages and Disadvantages\label{sec:adv-disadv-qc4qmd}}

The primary advantage of quantum simulation is exponential capacity in
representing system state with linear hardware growth. A quantum simulator
of $n$ elements embodies the full $2^n$-dimensional state space, using
only $n$ physical components. While memory and runtime of classical
computation grow exponentially with system size, quantum simulation of
local Hamiltonian dynamics requires resources growing only linearly (or
polynomially) with system size. This is the primitive reason why real-time
many-body dynamics is regarded as one of the most promising near-term
applications. Because the quantum simulator is a real quantum many-body
system under microscopic control, it gives access to quantities or processes
that are awkward or impossible in classical computation, such as real-time dynamics
far from equilibrium, entanglement growth, non-local string order parameters,
and finite-temperature states prepared by physical cooling rather than by
imaginary-time sampling. Analog experiments provide direct physical insight
because the target Hamiltonian is implemented natively through controllable
tunneling, interactions, confinement potentials, or long-range couplings.
The quantum simulator is thus best understood not merely as a faster computer
but as a programmable laboratory. Thus, quantum simulators routinely realize
experimental regimes that conventional laboratories do not offer. The single-particle
and single-shot measurement capability, which measures the full microscopic
configuration rather than a bulk average, is something neither a real
experiment nor a classical simulation provides. Indeed, the real experiment
cannot ``see'' individual electrons; meanwhile, the classical computation
cannot hold the quantum state exactly. This advantage is further extended
by tunability. For instance, Rydberg-atom arrays can be rearranged into
arbitrary lattice geometries with programmable interactions, enabling
the exploration of ordered phases, constrained spin models, and topological
spin-liquid signatures across hundreds of atoms. In this context, quantum
simulation converts some of open questions from computational problems
into experimental ones.

For the second advantage of quantum simulation, its applications span
many fields of science and technology in a way few technologies match
while it does not have to waite for the fault-tolerant era. Each target
of these applications is a strongly correlated system whose classical
treatment is either intractable or unacceptably approximate. Analog
quantum simulators circumvent the requirement for extensive error
correction because the target dynamics are native to the device and
they have already produced a string of results that stand as genuine
scientific contributions. On the digital quantum simulation side, the
transition toward error-corrected simulation has begun in earnest.
Although physical qubit counts now exceed a thousand on several
platforms while logical qubits (the currency of reliable simulation)
number in the tens, the development rate is steeply positive, and
industry roadmaps put chemistry-relevant digital quantum simulation
on a visible horizon. On the other hand, the speed at which the cost
model of quantum simulation is improving, independent of hardware. It
is worth noting five orders of magnitude improvement over past decade,
despite the underlying algorithm remaining unchanged. The gains came
from compilation strategy, Hamiltonian decomposition, and architecture
matching. Every such improvement multiplies the value of whatever
hardware exists. 

Having the above advantages, let us turn to disadvantages of quantum
computation for quantum molecular dynamics. The first disadvantage in
the present era is noise, as current quantum simulation devices operate
in the NISQ regime, that is tens to thousands
of physical qubits without full error correction, setting a hard ceiling
on circuit depth and simulation accuracy. In general, error accumulation
scales non-linearly with both qubit count and circuit depth, degrading
sampled distributions and imposing practical constraints on how expressive
a quantum circuit can be without error correction. For time-dependent and
interacting quantum systems, these limitations impose strict bounds on
achievable accuracy, particularly over long evolution times. In short, the
very feature that makes quantum simulators powerful also makes them fragile,
as large quantum systems are exquisitely sensitive to their environments.
Current leading platforms show physical error rates around $0.1\sim1\%$
per quantum gate, while efficient error correction wants physical rates
near or below $\sim0.01\%$, and textbook large-scale fault-tolerant
algorithms demand logical error rates of $10^{-10}\sim10^{-15}$. Error
mitigation can stretch NISQ accuracy but is statistical post-processing
rather than true correction, and it carries sampling costs that grow with
the noise level. Currently, logical qubits are real, demonstrated, and few.
A useful fault-tolerant simulation needs hundreds to thousands of them
executing millions of operations, several orders of magnitude beyond
anything running today. Until that gap closes, digital quantum simulation
inherits every hardware defect of quantum computing while analog simulation
sidesteps only some of them.
Quantum error correction (QEC) converts noisy physical qubits into reliable
logical qubits at a staggering exchange rate. QEC also imposes a second,
temporal bottleneck, that is syndrome decoding must run in real time, faster
than errors accumulate, across thousands of qubits. No one has yet executed
an end-to-end fault-tolerant algorithm on logical qubits. The disadvantage,
in one sentence, is that the quantum simulation of a molecule that matters
is, today, a machine of millions of qubits that does not exist.

The next disadvantage focuses on the limited programmability, algorithmic
errors, and hidden sampling costs in quantum computing. Analog quantum
simulators can only realize the restricted set of models natively supported
by the hardware, and even those computations are bounded by control,
calibration, and decoherence errors that are difficult to characterize.
This is because Hamiltonian engineering is itself an approximation of
the real system. On the other hand, by digital quantum simulation, Suzuki-Trotter
decomposition introduces systematic errors. This problem will quickly
become a limitation if we simulate larger problems aiming for higher
accuracy. Moreover, reducing these errors costs gates. For instance, a
$k$-th-order formula scales as roughly $n^{(1+1/k)}\cdot t^{(1+1/k)}\cdot\epsilon^{(-1/k)}$
in system size $n$, time $t$, and inverse accuracy $\epsilon$. Each Trotter
step demands multiple entangling gates per qubit, and on superconducting
hardware those non-local gates are the dominant noise source, forcing
uncomfortable trade-offs between algorithmic error (fewer, coarser steps)
and hardware error (shallower circuits). State preparation and measurement
add further overheads that rarely make headlines. In preparing a chemically
meaningful initial state, one requires a deep circuit. In extracting observables,
one requires statistical repetition, such as zero-noise extrapolation,
which may multiply the shot count further. Therefore, overhead of quantum
simulation is not only about qubits but also about gate counts, shot counts,
and wall-clock.

Finally, every hardware platform purchases some advantages by accepting
specific disadvantages, summarized in Table \ref{tab:qc-platform-1}.
Superconducting circuits offer the fastest gates (tens of nanoseconds)
and the most mature fabrication pipeline, but demand dilution refrigeration,
suffer crosstalk, and restrict connectivity to nearest neighbors on a
2D lattice. Trapped ions deliver the highest gate fidelities, superb
readout, and all-to-all connectivity, but gate times in the tens of
microseconds make them orders of magnitude slower in wall-clock, and
scaling beyond single chains faces ion-transport and crosstalk challenges.
Neutral atoms scale superbly, but their best two-qubit gate fidelities
($\sim99.5\%$) trail ions, preparation and atom rearrangement consume
most of each experimental cycle, and atoms are lost after measurement,
forcing full re-preparation. Photonic platforms bring room-temperature
operation and near-zero decoherence in transit, yet two-photon entangling
gates are probabilistic and photon loss dominates as circuits grow. Silicon
spin qubits leverage semiconductor manufacturing, but remain at small
qubit counts. Topological qubits promise intrinsic error protection and
radical overhead reduction, but no error-corrected logical computation
has been demonstrated on the approach. The dispersion of these trade-offs
matters for quantum simulation specifically because simulation workloads
are unusually sensitive to all of the parameters at once. The absence of
a dominant substrate means bet-hedging across platforms, which fragments
software, benchmarks, and expertise.

%------------------------
\subsection{Discussions\label{sec:mol-fiel}}

The advantages and disadvantages given in Section \ref{sec:adv-disadv-qc4qmd}
are not symmetric in time or in domain. First of all, where quantum
simulation already wins is the qualitative exploration of strongly
correlated quantum system. For insatnce, analog quantum simulation
platforms have delivered genuine discoveries, such as Hubbard-model
antiferromagnetism, many-body localization, Rydberg spin liquids, that
classical computation had not reached and in other regimes still have
not. Currently, the value of these quantum simulations is their insight
per experiment, but not floating-point calculations. The absence of
error correction requires physical robustness of the phenomena probed.
The second win of quantum simulation is instrumental, that quantum
simulators as controllable model systems for physics which is otherwise
inaccessible in other laboratory. Third, where the disadvantages dominate
is quantitative, say simulations at chemical accuracy, precisely the
applications with the clearest revenue logic. These demand thousands
of error-corrected logical qubits and billions of gates, and no such
machine exists. In between lies a contested middle ground, such as NISQ
utility experiments and quantum annealing. The balance is not that the
middle ground is empty, but that it is unstable.
As given in Table \ref{tab:qc-platform-1}, nearly every advantage of
quantum simulation has a paired disadvantage that is the same feature
viewed from the other side. The exponential state space that gives
simulators their power is what makes their outputs unverifiable. The
physical nativeness that frees analog devices from error correction is
what makes them unprogrammable. The quantum hardware sensitivity to its
environment is the source of its noise. This duality is why serious
assessments avoid both the hype register and the dismissal register.
Therefore, disadvantages of quantum simulation are engineering problems
with visible solutions, while limitations of its classical competitors
are complexity-theoretic hardness.

There exist three concrete stances in quantum computing. First, engagement through
the cloud is economically rational, where quantum simulation is available
at near-zero cost. Problem selection should follow the classical failure
modes. For instance, targets with sign problems, real-time dynamics, or
high-entanglement structure can offer durable headroom of quantum simulation,
whereas equilibrium properties of locally interacting low-dimensional
systems are exactly where classical computing will be a better choice
than quantum simulation. Third, claims should be held to the epistemic
standard arisen from classical computing, such as error characterization,
cross-platform inspections, and classical baselines. It is worth noting
what the disadvantages of quantum simulation are not. They do not include
a known fundamental obstruction. No theorem forbids scalable, error-corrected
quantum simulation. Below-threshold demonstrations have retired the deepest
skepticism. These disadvantages also do not include evidence that classical
computing methods will keep pace indefinitely. The rebuttals so far have
exploited specific low-entanglement structures of the chosen benchmark
experiments. Each successive claim has forced classical computing methods
to work measurably harder. Therefore, disadvantages of quantum simulation
can be seen as transient engineering and economic constraints, but are
qualitatively different from the permanent complexity barriers that motivated
Feynman in the first place.

%-----------------------------------------------------------------------------
% Conclusions
%-----------------------------------------------------------------------------
\section{Conclusions\label{sec:con}}

In this work, we explored the implementation possibility of quantum
simulation for quantum molecular dynamics, in particular for reaction
dynamics. Although several implementations have already reported through
quantum-classical mixed simulations
\cite{cle10:1155,geo14:153,gue19:045001,cao19:10856,oll20:260511,oll20:043140,oll21:4229,bha22:015004,zhu26:xxx},
there indeed exist several theoretical problems in this aspect. To
analyze these problems, we examined (1) the conjugacy relation between
quantum simulator and the target molecular system, (2) the wave function
correspondence in quantum algorithm and classical algorithm for
multi-dimensional dynamics, (3) problems arisen from real-valued
classical algorithms, and finally (4) geometric phase arisen from
the separation among the degrees of freedom (DOFs). As is well known,
the first and second points play fundamental roles in quantum simulation
of quantum many-body systems, and the third and fourth points are
theoretical issues that might introduce problems in classical and
quantum computing \cite{oll20:260511,oll20:043140,oll21:4229,zhu26:xxx,zha25:20397}.
In this work, we mainly focus on the third and
fourth points by analysis of the first two points by reviewing previously
reported quantum-classical mixed implementations of quantum
simulation. We also consider gauge freedom in high-dimensional quantum
molecular dynamics that has been introduced recently, and then discuss
possibility of advantages and disadvantages of quantum simulation
for molecular reaction dynamics.

%--------------------------------------------------------------------------
% Supplementary Material
%-------------------------------------------------------------------------
\section*{Supplementary Material}

The Supporting Information file is available free of charge at
https://www.doi.org/XXXX. 

%--------------------------------------------
% Declaration of interests
%----------------------------------------------
\section*{Interest Statement}

The authors declare that they have no known competing financial interests
or personal relationships that could have appeared to influence the work
reported in this paper.

%----------------------------------------------------------------
% Data availability
%-----------------------------------------------------------------
\section*{Data Availability}

All data have been reported in this work. The data supporting this article
have been included as part of the Supplementary Information.

%--------------------------------------------------------------------
% Acknowledgements
%--------------------------------------------------------------------
\section*{Acknowledgments}

The financial supports of National Natural Science Foundation of China
(Grant No. 22273074) and Fundamental Research Funds for the Central
Universities (Grant Nos. 2025JGZY34 and 2025KCW017) are gratefully
acknowledged. The authors are also grateful to anonymous reviewers for
their thoughtful suggestions.

%------------------------------------------------
% Tables
%------------------------------------------------
%----------------Compare geom--------------------
\clearpage
\begin{sidewaystable}
%\begin{table}[h!]
 \caption{
Comparisons of the concepts of the geometric phase effects in the nuclear
and electronic properties, together with these concepts associated with
electromagnetic field. The second column provides geometric concepts
that will be compared, including the Berry connection, Berry curvature,
geometric phase, and Chern number. The third and fourth columns list
physical insight of these concepts in the nuclear and electronic
properties, respectively. The fifth column lists the electromagnetic
quantities or electromagnetic properties associated with the geometric
features. The rightmost column gives remarks of these physical quantities
or geometric properties.}
 \begin{tabular}{lcllllllllr}
  \hline
No. &~~& Characteristic &~~& Nuclear &~~& Electron &~~& Electromagnetic Field &~~& Remarks \\
\hline
1 && Berry connection
&~~& $A(\mathbi{Q})=\mi\langle\Psi\vert\partial_{\mathbi{Q}}\vert\Psi\rangle_{\mathbi{q}}$
&~~& $\tau_{ii}^{\mathrm{elec}}(\mathbi{y})=
\langle\phi_i\vert\nabla_{\mathbi{y}}\vert\phi_i\rangle_{\mathbi{x}}$
\footnote{If the $j$-th electronic state $\phi_j(\mathbi{x},\mathbi{y})$
depends on electronic cooridnates $\mathbi{x}$ and parameters $\mathbi{y}$,
then $\tau_{ij}^{\mathrm{elec}}(\mathbi{y})$ is elements of the electronic
non-adiabatic coupling matrix.\label{foot:elec-connection}}
&~~& analogous to the vector potential
&~~& It is a gauge-dependent object   \\
&& && && &&  &~~& encoding system's dependence  \\
&& && && &&  &~~& on the parameters space
\footnote{The parameter space, a key notion in differential geometry,
represents the set of configurations available to the system, while
its dynamics is governed by the EOMs.\label{foot:geom-em-field}}. \\
2&& Berry curvature  &~~& curl of $A(\mathbi{Q})$
           &~~& curl of $\tau_{ii}^{\mathrm{elec}}(\mathbi{y})$
\footnote{The Berry curvature of the $i$-th electronic state is
$\Omega_i(\mathbi{y})=\nabla_{\mathbi{y}}\times\tau_{ii}^{\mathrm{elec}}(\mathbi{y})$.
See also Footnote \textsuperscript{\ref{foot:elec-connection}}.}
           &~~& effective magnetic field in 
           &~~& It is a local geometric property of \\
&& && && && parameter space \textsuperscript{\ref{foot:geom-em-field}}
           &~~& parameter space encoding the  \\
&& && && &&  &~~& states' anholonomy under  \\
&& && && &&  &~~& adiabatic parameter variations.  \\ 
3&& Berry phase &~~& in nuclear properties &~~& in electronic properties
           &~~& geometric propertities the  &~~& It is a geometric properties of \\ 
&& && && && of electromagnetic field 
\footnote{For example, the Berry phase effects of the electromagnetic
field can be found in classical polarization dynamics and quantum
photonic system.}  &~~& wave eigen-functions, such as \\
&& && && &&    &~~& wave funciton and field function. \\
4&& Chern number \footnote{The Chern number is equal to number of 
critical points in the parameter space
\textsuperscript{\ref{foot:geom-em-field}}, in which the system is evolving.}
&& number of pathological  
&~~& number of conical &~~& global properties of the &~~& It is a topological invariant  \\
&& && points in environment && intersection points && electromagnetic modes
\footnote{In general, the electromagnetic modes are defined for photons,
or more pricisely bosons, such as photon bands in periodic structures.}
&~~& classifing the global topology of  \\
&& && && &&  &~~& the system’s eigenstate bundle  \\ 
\hline
\end{tabular}
\label{tab:geom-comp}
\end{sidewaystable}

%-table-2----------------------------------
\clearpage
\begin{table}
\caption{Leading quantum computing device platforms that have been
already established. The first and secod columns give platforms and
developers, respectively. The third column gives number of qubits. The
other columns give advantages and limitations of each platform.}
\label{tab:qc-platform-1}
\end{table}
%----------------------
\clearpage
\begin{sidewaystable}
\begin{tabular}{lllll}
\hline
Platform & Institution/Company & Qubit Count & Strengths & Limitations \\
\hline
\multicolumn{5}{l}{{\it Superconducting}} \\
Condor & IBM & 1121 & High qubit count; mature ecosystem & Requires cryogenic cooling; limited connectivity \\
Eagle & IBM & 127 & Accessible via cloud; broad software support & Noise and error rates limit depth \\
Heron & IBM & 133 & Improved error rates & \\
Willow & Google & 105 & 0.143\% error rate per cycle & Limited public access \\
Sycamore & Google & 53 & Demonstrated quantum supremacy & \\
Zuchongzhi 3.0 & USTC & -- & Competing with Google/IBM &  \\
\hline
\multicolumn{5}{l}{{\it Trapped Ion}} \\
Quantinuum H1-2 & Quantinuum & 20--56 & High fidelity; all-to-all connectivity & Slow gate speeds; challenging to scale \\
IonQ Aria & IonQ & 29--32 & Long coherence times & Cost; complex laser systems \\
UMD QEC System & UMD & $\sim$228 & Quantum error correction focus & Research only \\
%\hline
%\end{tabular}
%\label{tab:qc-platform-0}
%\end{sidewaystable}
%
%%------table-3----
%\clearpage
%\begin{sidewaystable}
%\caption{Leading Quantum Computing Platforms (2026)}
%\begin{tabular}{lllll}
%\hline
%Platform & Institution/Company & Qubit Count & Strengths & Limitations \\
\hline
\multicolumn{5}{l}{{\it Neutral Atom}} \\
QuEra QPU & Harvard/MIT/QuEra & $\sim$280 & Large qubit count; flexible connectivity & Early-stage error correction; laser noise \\
PASQAL & PASQAL & $\sim$100 & Configurable geometries; analog and digital modes & Less mature than superconducting \\
ETH QEC Array & ETH Zurich/PSI & $\sim$200 & Research platform & Academic access only \\
\hline
\multicolumn{5}{l}{{\it Photonic}} \\
Alibaba/Xanadu & Alibaba/Xanadu & $\sim$24 & Room temperature operation & Photon loss limits scaling \\
\hline
\multicolumn{5}{l}{{\it Quantum Annealing}} \\
Advantage & D-Wave & 5000+ & Largest qubit count; solves optimization problems & Limited to optimization; not universal \\
\hline
\multicolumn{5}{l}{{\it Spin Qubit}} \\
Tunnel Falls & Intel & -- & Compatible with semiconductor fabrication & Early-stage technology \\
\hline
\multicolumn{5}{l}{{\it Topological Qubit}} \\
Majorana-1 & Microsoft & $\sim$67--70 & Potentially fault-tolerant & Proprietary; unproven \\
\hline
\end{tabular}
%\label{tab:qc-platform-1}
\end{sidewaystable}

%-------------------------------------------------
% FIGURES
%-------------------------------------------------
\clearpage
\section*{Figure Captions}

\figcaption{fig:vqe-for-quan-chem}{
Simple illustration of the VQE algorithm for electronic-structure theory
in quantum computing. In VQE, quantum computer is used to prepare a set
of parametrized quantum states as shown in green box. The classical computer 
then takes the individual estimates of the wave function parameters and
averages them to compute a single value of each parameter as shown in
yellow box. This cost
function value is fed into an optimization routine, which produces an
updated set of parameters as input for the quantum circuit in the next
optimization loop. This procedure is repeated until the energy converges.
Recent and future efforts are to improve each VQE component as well as
replace certain ``classical'' subroutines with quantum counterparts to
further leverage the capabilities of quantum computers. Reprinted with
permission from Reference \cite{cao19:10856}. Copyright 2019 American
Chemical Society.
}

\figcaption{fig:quan-eom-chem}{
Illuatration of the quantum equation of motion (qEOM) algorithm. First,
the Jordan-Wigner transformation is used to map the commutators for
computing $\boldsymbol{\mathcal{M}}$ and $\boldsymbol{\mathcal{V}}$.
These commutators are originally expressed in terms of the Fermionic
creation and annihilation operators into the qubits space, and are
then evaluated using the ground-state wave function prepared in the
quantum hardware from a VQE calculation to obtain matrices in the
EOM. From these measurements the secular equation is constructed.
Its eigenvalues are then classically solved to obtain the first
excitation (and corresponding de-excitation) energies. Reprinted with
permission from Reference \cite{oll20:043140}. Copyright 2020 American
Physical Society.
}

\figcaption{fig:prop-quan-simul}{
Quantum circuit for the time evolution of the time-dependent wave
function based on the Hamiltonian of the Marcus spin-boson model,
$\hat{H}=\hat{K}\otimes I+V_0\otimes\vert0\rangle\langle0\vert+
V_1\otimes\vert1\rangle\langle1\vert+C\otimes\sigma_x$. The $K$, $V_i$,
and $C$ blocks represent the time evolution operators for the kinetic
energy, $i$-th potential with $i=0,1$, and coupling terms, respectively.
The dynamical bosonic degrees of freedom in the model are represented
as a wave packet evolving under the action of a harmonic oscillator
Hamiltonian, where the displacement operators are coupled with the
$\sigma_x$ operator of the spin. Reprinted with permission from
Reference \cite{oll20:260511}. Copyright 2020 American Physical
Society.
}

\figcaption{fig:marcus-quancomp}{
Quantum simulation results on the Marcus spin-boson model given in
Equation \eqref{eq:marcus-model-ham-01}, including (a) the time-dependent
population fraction $P_0$ in the product well, (b) linear fitting of
the first ten steps of $P_0$ to approximate the rate constants, (c)
rate constants $k$ as function of the offset. In subfigure (a), results
obtained with the algorithm \cite{oll20:260511} given in Section \ref{sec:implem-appli-qc}
in classical simulations are given by dots, while the exact evolutions
with the reference coupling and the approximate coupling are given by
solid lines and dashed lines, respectively. In subfigures (a) and (b),
colors represent different offset values between $V_0$ and $V_1$. In
subfigure (c), results obtained with the algorithm \cite{oll20:260511}
in Section \ref{sec:implem-appli-qc} are given by dots, while the
exact evolution with the reference coupling is given by crosses. The
Marcus rates are shown as a dashed line for a qualitative comparison.
The colored stickers label different charge transfer regions (A, normal
regime; B, at reorganization energy; C, inverted region). Reprinted
with permission from Reference \cite{oll20:260511}. Copyright 2020
American Physical Society.
}

\figcaption{fig:nocl-quan-circuit}{
Illustration of quantum circuits for (a) a single propagation step $\Delta t$ and
(b) the Hadamard test used to extract the autocorrelation function.
In subfigure (a), the circuit implements the second-order split-operator
method, utilizing the quantum Fourier transform (QFT) and Legendre-DVR
transformation ($U_{\theta}$) to switch between the DVR and FBR. The
nonunitary complex absorbing potential is implemented via dilation
unitary matrix $U_{\mathrm{dil}}$, which requires an additional ancilla
qubit $\vert0\rangle_a$. In subfigure (b), the quantum circuit for the
Hadamard test used to extract the autocorrelation function has a control
ancilla qubit $\vert0\rangle_a$ that governs the application of total
time evolution operator. Reprinted with permission from Reference
\cite{zhu26:xxx}. Copyright 2026 American Chemical Society.
}

\figcaption{fig:nocl-spectrum}{
Photodissociation cross section comparison of (a) spectrum from the ideal
$2^{10}$-shot simulation with (b) spectrum from the GuadalupeV2 noise model
simulation. The Fourier transform effectively smooths the hardware-induced
noise, preserving the core physically meaningful features. The yellow solid
and the blue dash lines are results from quantum and classical simulations,
respectively. Reprinted with permission from Reference \cite{zhu26:xxx}.
Copyright 2026 American Chemical Society.
}

%%----------fig-1---------------
\clearpage
 \begin{figure}[h!]
  \centering
   \includegraphics[width=18cm]{./vqe-alg.pdf}
    \caption{\figfoot}
     \label{fig:vqe-for-quan-chem}
      \end{figure}

%%----------fig-2---------------
\clearpage
 \begin{figure}[h!]
  \centering
   \includegraphics[width=18cm]{./qeom.pdf}
    \caption{\figfoot}
     \label{fig:quan-eom-chem}
      \end{figure}
      
%%----------fig-3---------------
\clearpage
      
%-----------fig-4-------------
\clearpage
 \begin{figure}[h!]
  \centering
   \includegraphics[width=18cm]{./marcus-quancomp.pdf}
    \caption{\figfoot}
     \label{fig:marcus-quancomp}
      \end{figure}

%------fig.-5--------------
\clearpage
 \begin{figure}[h!]
  \centering
   \includegraphics[width=18cm]{./nocl-quan-circuit.pdf}
    \caption{\figfoot}
     \label{fig:nocl-quan-circuit}
      \end{figure}

%------fig.-6--------------
\clearpage
 \begin{figure}[h!]
  \centering
   \includegraphics[width=18cm]{./nocl-spectrum.pdf}
    \caption{\figfoot}
     \label{fig:nocl-spectrum}
      \end{figure}

%-----------------------------------------------------
% References
%------------------------------------------------------


\clearpage

\begin{thebibliography}{10}
\ifnum\language=1 \def\biband{und}\else\def\biband{and}\fi

\bibitem{cle10:1155}
{\rm A.~A. Clerk, M.~H. Devoret, S.~M. Girvin, F.~Marquardt,  \biband~ R.~J.
  Schoelkopf}.
\newblock Introduction to quantum noise, measurement, and amplification.
\newblock {\em Rev. Mod. Phys. \bf 82\/} (2010), 1155--1208.

\bibitem{geo14:153}
{\rm I.~M. Georgescu, S.~Ashhab,  \biband~ F.~Nori}.
\newblock Quantum simulation.
\newblock {\em Rev. Mod. Phys. \bf 86\/} (2014), 153--185.

\bibitem{gue19:045001}
{\rm D.~Gu\'ery-Odelin, A.~Ruschhaupt, A.~Kiely, E.~Torrontegui,
  S.~Mart\'{\i}nez-Garaot,  \biband~ J.~G. Muga}.
\newblock Shortcuts to adiabaticity: Concepts, methods, and applications.
\newblock {\em Rev. Mod. Phys. \bf 91\/} (2019), 045001.

\bibitem{cao19:10856}
{\rm Y.~Cao, J.~Romero, J.~P. Olson, M.~Degroote, P.~D. Johnson,
  M.~Kieferov{\'a}, I.~D. Kivlichan, T.~Menke, B.~Peropadre, N.~P.~D. Sawaya,
  S.~Sim, L.~Veis,  \biband~ A.~Aspuru-Guzik}.
\newblock Quantum chemistry in the age of quantum computing.
\newblock {\em Chem.\ Rev. \bf 119\/} (2019), 10856--10915.

\bibitem{oll20:260511}
{\rm P.~J. Ollitrault, G.~Mazzola,  \biband~ I.~Tavernelli}.
\newblock Nonadiabatic molecular quantum dynamics with quantum computers.
\newblock {\em Phys. Rev. Lett. \bf 125\/} (2020), 260511.

\bibitem{oll20:043140}
{\rm P.~J. Ollitrault, A.~Kandala, C.-F. Chen, P.~K. Barkoutsos, A.~Mezzacapo,
  M.~Pistoia, S.~Sheldon, S.~Woerner, J.~M. Gambetta,  \biband~ I.~Tavernelli}.
\newblock Quantum equation of motion for computing molecular excitation
  energies on a noisy quantum processor.
\newblock {\em Phys. Rev. Res. \bf 2\/} (2020), 043140.

\bibitem{oll21:4229}
{\rm P.~J. Ollitrault, A.~Miessen,  \biband~ I.~Tavernelli}.
\newblock Molecular quantum dynamics: A quantum computing perspective.
\newblock {\em Acc.\ Chem.\ Res. \bf 54\/} (2021), 4229--4238.

\bibitem{bha22:015004}
{\rm K.~Bharti, A.~Cervera-Lierta, T.~H. Kyaw, T.~Haug, S.~Alperin-Lea,
  A.~Anand, M.~Degroote, H.~Heimonen, J.~S. Kottmann, T.~Menke, W.-K. Mok,
  S.~Sim, L.-C. Kwek,  \biband~ A.~Aspuru-Guzik}.
\newblock Noisy intermediate-scale quantum algorithms.
\newblock {\em Rev. Mod. Phys. \bf 94\/} (2022), 015004.

\bibitem{zhu26:xxx}
{\rm Z.~Zhuang, C.~Yang, Y.~Wang, D.~H. Zhang,  \biband~ B.~Zhao}.
\newblock Quantum mechanical studies of photodissociation dynamics on quantum
  computers.
\newblock {\em J.~Phys.\ Chem.\ Lett.\/} (2026).

\bibitem{wan03:1289}
{\rm H.~Wang \biband~ M.~Thoss}.
\newblock Multilayer formulation of the multiconfiguration time-dependent
  {H}artree theory.
\newblock {\em J.~Chem.\ Phys. \bf 119\/} (2003), 1289--1299.

\bibitem{man08:164116}
{\rm U.~Manthe}.
\newblock A multilayer multiconfigurational time-dependent {H}artree approach
  for quantum dynamics on general potential energy surfaces.
\newblock {\em J.~Chem.\ Phys. \bf 128\/} (2008), 164116.

\bibitem{ven11:044135}
{\rm O.~Vendrell \biband~ H.-D. Meyer}.
\newblock {Multilayer multiconfiguration time-dependent Hartree method:
  Implementation and applications to a Henon-Heiles Hamiltonian and to
  pyrazine}.
\newblock {\em J.~Chem.\ Phys. \bf 134\/} (2011), 044135.

\bibitem{wan15:7951}
{\rm H.~Wang}.
\newblock {Multilayer Multiconfiguration Time-Dependent Hartree Theory}.
\newblock {\em J.~Phys.\ Chem.~A \bf 119\/} (2015), 7951.

\bibitem{zha25:20397}
{\rm X.~Zhang \biband~ Q.~Meng}.
\newblock A hierarchical wavepacket propagation framework via ml-mctdh for
  molecular reaction dynamics.
\newblock {\em Phys.\ Chem.\ Chem.\ Phys. \bf 27\/} (2025), 20397--20420.

\bibitem{wei20:060406}
{\rm M.~Weilenmann \biband~ R.~Colbeck}.
\newblock Self-testing of physical theories, or, is quantum theory optimal with
  respect to some information-processing task?
\newblock {\em Phys. Rev. Lett. \bf 125\/} (2020), 060406.

\bibitem{ren21:625}
{\rm M.-O. Renou, D.~Trillo, M.~Weilenmann, T.~P. Le, A.~Tavakoli, N.~Gisin,
  A.~Ac{\'i}n,  \biband~ M.~Navascu{\'e}s}.
\newblock Quantum theory based on real numbers can be experimentally falsified.
\newblock {\em Nature \bf 600\/} (2021), 625--629.

\bibitem{wu22:140401}
{\rm D.~Wu, Y.-F. Jiang, X.-M. Gu, L.~Huang, B.~Bai, Q.-C. Sun, X.~Zhang, S.-Q.
  Gong, Y.~Mao, H.-S. Zhong, M.-C. Chen, J.~Zhang, Q.~Zhang, C.-Y. Lu,
  \biband~ J.-W. Pan}.
\newblock Experimental refutation of real-valued quantum mechanics under strict
  locality conditions.
\newblock {\em Phys. Rev. Lett. \bf 129\/} (2022), 140401.

\bibitem{li22:040402}
{\rm Z.-D. Li, Y.-L. Mao, M.~Weilenmann, A.~Tavakoli, H.~Chen, L.~Feng, S.-J.
  Yang, M.-O. Renou, D.~Trillo, T.~P. Le, N.~Gisin, A.~Ac\'{\i}n,
  M.~Navascu\'es, Z.~Wang,  \biband~ J.~Fan}.
\newblock Testing real quantum theory in an optical quantum network.
\newblock {\em Phys. Rev. Lett. \bf 128\/} (2022), 040402.

\bibitem{bar26:240202}
{\rm P.~{Barrios Hita}, A.~Trushechkin, H.~Kampermann, M.~Epping,  \biband~
  D.~Bru{\ss}}.
\newblock Quantum mechanics based on real numbers: A consistent description.
\newblock {\em Phys. Rev. Lett. \bf 136\/} (2026), 240202.

\bibitem{men22:16415}
{\rm Q.~Meng, J.~Chen, J.~Ma, X.~Zhang,  \biband~ J.~Chen}.
\newblock Adiabatic models for the quantum dynamics of surface scattering with
  lattice effects.
\newblock {\em Phys.\ Chem.\ Chem.\ Phys. \bf 24\/} (2022), 16415--16436.

\bibitem{men15:164310}
{\rm Q.~Meng \biband~ H.-D. Meyer}.
\newblock {Expansion Hamiltonian model for a diatomic molecule adsorbed on a
  surface: Vibrational states of the CO/Cu(100) system including surface
  vibrations}.
\newblock {\em J.~Chem.\ Phys. \bf 143\/} (2015), 164310.

\bibitem{men17:184305}
{\rm Q.~Meng \biband~ H.-D. Meyer}.
\newblock {Lattice effects of surface cell: Multilayer multiconfiguration
  time-dependent Hartree study on surface scattering of CO/Cu(100)}.
\newblock {\em J.~Chem.\ Phys. \bf 146\/} (2017), 184305.

\bibitem{men21:2702}
{\rm Q.~Meng, M.~Schr{\"o}der,  \biband~ H.-D. Meyer}.
\newblock High-dimensional quantum dynamics study on excitation-specific
  surface scattering including lattice effects of a five-atom surface cell.
\newblock {\em J.~Chem.\ Theory Comput. \bf 17\/} (2021), 2702--2713.

\bibitem{ha25:19423}
{\rm J.-K. Ha \biband~ R.~J. MacDonell}.
\newblock Analog quantum simulation of coupled electron-nuclear dynamics in
  molecules.
\newblock {\em Chem. Sci. \bf 16\/} (2025), 19423--19435.

\bibitem{pan12:777}
{\rm J.-W. Pan, Z.-B. Chen, C.-Y. Lu, H.~Weinfurter, A.~Zeilinger,  \biband~
  M.~$\dot{\mathrm{Z}}$ukowski}.
\newblock Multiphoton entanglement and interferometry.
\newblock {\em Rev.\ Mod.\ Phys. \bf 84\/} (2012), 777--838.

\bibitem{die82:271}
{\rm D.~Dieks}.
\newblock Communication by {EPR} devices.
\newblock {\em Phys. Lett. A \bf 92\/} (1982), 271--272.

\bibitem{hen85:1}
{\rm M.~Henneaux}.
\newblock Hamiltonian form of the path integral for theories with a gauge
  freedom.
\newblock {\em Phys. Rep. \bf 126\/} (1985), 1--66.

\bibitem{fro93:733}
{\rm J.~Fr\"ohlich \biband~ U.~M. Studer}.
\newblock Gauge invariance and current algebra in nonrelativistic many-body
  theory.
\newblock {\em Rev. Mod. Phys. \bf 65\/} (1993), 733--802.

\bibitem{lit97:213}
{\rm R.~G. Littlejohn \biband~ M.~Reinsch}.
\newblock Gauge fields in the separation of rotations andinternal motions in
  the $n$-body problem.
\newblock {\em Rev. Mod. Phys. \bf 69\/} (1997), 213--276.

\bibitem{raw21:24154}
{\rm J.~I. Rawlinson, C.~F{\'a}bri,  \biband~ A.~G. Cs{\'a}sz{\'a}r}.
\newblock The rovibrational {A}haronov-{B}ohm effect.
\newblock {\em Phys. Chem. Chem. Phys. \bf 23\/} (2021), 24154--24164.

\bibitem{aha91:818}
{\rm Y.~Aharonov, E.~Ben-Reuven, S.~Popescu,  \biband~ D.~Rohrlich}.
\newblock Born-oppenheimer revisited.
\newblock {\em Nucl. Phys. B \bf 350\/} (1991), 818--830.

\bibitem{boh92:53}
{\rm A.~Bohm, B.~Kendrick,  \biband~ M.~E. Loewe}.
\newblock The berry phase in molecular physics.
\newblock {\em Int.\ J.\ Quant.\ Chem. \bf 41\/} (1992), 53--75.

\bibitem{boh91:1206}
{\rm A.~Bohm, L.~J. Boya,  \biband~ B.~Kendrick}.
\newblock Derivation of the geometrical phase.
\newblock {\em Phys. Rev. A \bf 43\/} (1991), 1206--1210.

\bibitem{boh92:977}
{\rm A.~Bohm, B.~Kendrick, M.~E. Loewe,  \biband~ L.~J. Boya}.
\newblock The berry connection and born–oppenheimer method.
\newblock {\em J. Math. Phys. \bf 33\/} (1992), 977--989.

\bibitem{ken97:2431}
{\rm B.~Kendrick}.
\newblock Geometric phase effects in the vibrational spectrum of
  ${\mathrm{na}}_{3}(\mathit{X})$.
\newblock {\em Phys.\ Rev.\ Lett. \bf 79\/} (1997), 2431--2434.

\bibitem{ken03:6739}
{\rm B.~K. Kendrick}.
\newblock Geometric phase effects in chemical reaction dynamics and molecular
  spectra.
\newblock {\em J.~Phys.\ Chem.~A \bf 107\/} (2003), 6739--6756.

\bibitem{sim83:2167}
{\rm B.~Simon}.
\newblock Holonomy, the quantum adiabatic theorem, and berry's phase.
\newblock {\em Phys.\ Rev.\ Lett. \bf 51\/} (1983), 2167--2170.

\bibitem{sim88:1725}
{\rm R.~Simon \biband~ N.~Kumar}.
\newblock A note on the berry phase for systems having one degree of freedom.
\newblock {\em J. Phys. A: Math. Gen. \bf 21\/} (1988), 1725.

\bibitem{avr89:595}
{\rm J.~E. Avron, L.~Sadun, J.~Segert,  \biband~ B.~Simon}.
\newblock Chern numbers, quaternions, and berry's phases in fermi systems.
\newblock {\em Commun. Math. Phys. \bf 124\/} (1989), 595--627.

\bibitem{aga90:6924}
{\rm G.~S. Agarwal \biband~ R.~Simon}.
\newblock Berry phase, interference of light beams, and the hannay angle.
\newblock {\em Phys.\ Rev.~A \bf 42\/} (1990), 6924--6927.

\bibitem{ana92:307}
{\rm J.~Anandan}.
\newblock The geometric phase.
\newblock {\em Nature \bf 360\/} (1992), 307--313.

\bibitem{mea92:51}
{\rm C.~A. Mead}.
\newblock The geometric phase in molecular systems.
\newblock {\em Rev.\ Mod.\ Phys. \bf 64\/} (1992), 51--85.

\bibitem{mar22:206002}
{\rm R.~Martinazzo \biband~ I.~Burghardt}.
\newblock Quantum dynamics with electronic friction.
\newblock {\em Phys.\ Rev.\ Lett. \bf 128\/} (2022), 206002.

\bibitem{mar24:10416}
{\rm R.~Martinazzo \biband~ I.~Burghardt}.
\newblock Emergence of the molecular geometric phase from exact
  electron–nuclear dynamics.
\newblock {\em J.~Phys.\ Chem.\ Lett. \bf 15\/} (2024), 10416--10424.

\bibitem{mar24:243002}
{\rm R.~Martinazzo \biband~ I.~Burghardt}.
\newblock Dynamics of the molecular geometric phase.
\newblock {\em Phys.\ Rev.\ Lett. \bf 132\/} (2024), 243002.

\bibitem{wil84:2111}
{\rm F.~Wilczek \biband~ A.~Zee}.
\newblock Appearance of gauge structure in simple dynamical systems.
\newblock {\em Phys.\ Rev.\ Lett. \bf 52\/} (1984), 2111--2114.

\bibitem{fuj01:75}
{\rm K.~Fujii}.
\newblock Mathematical foundations of holonomic quantum computer.
\newblock {\em Rep. Math. Phys. \bf 48\/} (2001), 75--82.

\bibitem{zha23:1}
{\rm J.~Zhang, T.~H. Kyaw, S.~Filipp, L.-C. Kwek, E.~Sj{\"o}qvist,  \biband~
  D.~Tong}.
\newblock Geometric and holonomic quantum computation.
\newblock {\em Phys. Rep. \bf 1027\/} (2023), 1--53.

\bibitem{som16:062202}
{\rm R.~D. Somma}.
\newblock A {Trotter-Suzuki} approximation for {L}ie groups with applications
  to {H}amiltonian simulation.
\newblock {\em J. Math. Phys. \bf 57\/} (2016), 062202.

\bibitem{hat05:37}
{\rm N.~Hatano \biband~ M.~Suzuki}.
\newblock {\em Finding Exponential Product Formulas of Higher Orders}.
\newblock Springer Berlin Heidelberg, Berlin, Heidelberg, 2005, pp.~37--68.

\bibitem{jav24:2405}
{\rm A.~Javadi-Abhari, M.~Treinish, K.~Krsulich, C.~J. Wood, J.~Lishman,
  J.~Gacon, S.~Martiel, P.~D. Nation, L.~S. Bishop, A.~W. Cross, B.~R. Johnson,
   \biband~ J.~M. Gambetta}.
\newblock Quantum computing with {Qiskit}, 2024.

\bibitem{pat25:2508}
{\rm P.~Pathak, K.~Tarakeshwar, S.~S. Ali, S.~Devendrababu,  \biband~
  A.~Ganesan}.
\newblock The evolution of {IBM}'s quantum information software kit (qiskit): A
  review of its applications, 2025.

\end{thebibliography}
\end{document}